\documentclass
[aps,amsfonts,amssymb,amsmath,floats,superscriptaddress,prc,showpacs,%
twocolumn,nofootinbib,showkeys]{revtex4}

\newcommand{\eq}{Eq.~}

\newcommand{\refer}{Ref.~}
\newcommand{\refers}{Refs.~}
\newcommand{\figu}{Fig.~}
\newcommand{\figus}{Figs.~}
\newcommand{\tab}{Tab.~}
\newcommand{\sect}{Sec.~}

\newcommand{\footnoteb}[1]{$^($\footnote{#1}$^)$}

\usepackage{multirow}
\usepackage{longtable}
\usepackage{epsfig}
\usepackage{float}
\usepackage[english]{babel}
\usepackage{array}
\usepackage{graphics}
\usepackage[roman]{sublabel}

\newcommand{\trho}{{\tilde\rho}}

\begin{document}

\title{Halo phenomenon in finite many-fermion systems. \\
Atom-positron complexes and large-scale study of atomic nuclei}
\author{V. Rotival}
\email{vincent.rotival@polytechnique.org} \affiliation{DPTA/Service
de Physique Nucl\'eaire - CEA/DAM \^Ile-de-France - BP12 - 91680
Bruy\`eres-le-Ch\^atel, France}\affiliation{National Superconducting
Cyclotron Laboratory, 1 Cyclotron Laboratory, East Lansing, MI
48824, USA}
\author{K. Bennaceur}
\email{bennaceur@ipnl.in2p3.fr}
\affiliation{Universit{\'e} de Lyon, F-69003 Lyon, France;
             Institut de Physique Nucl{\'e}aire de Lyon,
             CNRS/IN2P3, Universit{\'e} Lyon 1,
             F-69622 Villeurbanne, France}
\affiliation{CEA, Centre de Saclay, IRFU/Service de Physique Nucl{\'e}aire, F-91191 Gif-sur-Yvette, France}
\author{T. Duguet}
\email{thomas.duguet@cea.fr}
\affiliation{CEA, Centre de Saclay, IRFU/Service de Physique Nucl{\'e}aire, F-91191 Gif-sur-Yvette, France}
\affiliation{National Superconducting
Cyclotron Laboratory, 1 Cyclotron Laboratory, East Lansing, MI
48824, USA} \affiliation{Department of Physics and Astronomy,
Michigan State University, East Lansing, MI 48824, USA}

\date{\today}

\begin{abstract}
The analysis method proposed in Ref.~\cite{rotival07a} is applied to characterize halo properties in finite many-fermion systems. First, the versatility of the method is highlighted by applying it to light and
medium-mass nuclei as well as to atom-positron and ion-positronium complexes. Second, the dependence of nuclear halo properties on the
characteristics of the energy density functional used in self-consistent
Hartree-Fock-Bogoliubov calculations is studied. We focus in particular on the influence of (i) the scheme used to regularize/renormalize the ultra-violet divergence of the local
pairing functional, (ii) the angular-momentum cutoff in the single-particle basis, as well as (iii) the isoscalar effective
mass, (iv) saturation density and (v) tensor terms characterizing the particle-hole part of the energy
functional. It is found that (a) the low-density behavior of the pairing functional and the
regularization/renormalization scheme must be chosen coherently and with care to provide meaningful predictions, (b) the impact of
pairing correlations on halo properties is significant and is the result of two competing effects, (c)
the detailed characteristics of the pairing functional has however only little importance, (d) halo properties
depend significantly on any ingredient of the energy density functional that influences the location of
single-particle levels; i.e. the effective mass, the tensor terms and the saturation density of nuclear matter. The latter dependencies give
insights to how experimental data on medium-mass drip-line nuclei can be used in the distant future to
constrain some characteristics of the nuclear energy density functional. Last but not least, large scale
predictions of halos among all spherical even-even nuclei are performed using specific sets of particle-hole and
particle-particle energy functionals. It is shown that halos in the ground state of medium-mass nuclei will only
be found at the very limit of neutron stability and for a limited number of elements.
\end{abstract}

\maketitle

\section{Introduction}
\label{sec:intro}

The formation of halos is a quantum phenomenon caused by the possibility for non-classical systems to expand in the classically forbidden region~\cite{hansen87,hansen95,tanihata96,tanihata03,jensen04}. Indeed, weakly-bound systems
can extend well beyond the classically-allowed region, as has been theoretically predicted or experimentally
observed for molecules (\mbox{$^3$He-$^3$He-$^{39}$K}~\cite{li06},
\mbox{$^4$He$_2$}~\cite{schollkopf96,nielsen98,grisenti00},  \mbox{$^3$He$^4$He$_2$}~\cite{bressanini02}...),
atom-positron complexes (\mbox{$e^+$Be}, \mbox{PsLi$^+$}, \mbox{PsHe$^+$}...)~\cite{mitroy05} and hypernuclei
(\mbox{$^3_\Lambda$H})~\cite{cobis97}. In nuclear physics, where the study of halos was initiated,
efforts are still devoted to reach a better understanding of the structure and reaction properties of such exotic
systems. For instance, the existence of halos in borromean systems or excited states of mirror nuclei still raises
questions~\cite{chen05}. In light nuclei, it was found that a cluster picture is at play, and known halo systems
are accurately described by two-~\cite{fedorov94,nunes96a} or three-body~\cite{zhukov93,fedorov94,nunes96b,bang96}
models, where one or two nucleons evolve around a tightly-bound core. This leads to a classification in terms of
one-nucleon halos (\mbox{$^{11}$Be}~\cite{tanihata88,fukuda91,zahar93},
\mbox{$^{19}$C}~\cite{bazin95,kanungo00}, \mbox{$^{17}$Ne}~\cite{kanungo03,jeppesen04}...) and two-nucleon
halos (\mbox{$^6$He}~\cite{zhukov93}, \mbox{$^{11}$Li}~\cite{tanihata85a,tanihata85b}...).

Since the discovery of the anomalous cross-section of \mbox{$^{11}$Li}~\cite{tanihata85a,tanihata85b}, one of the
compelling questions relates to the existence of a mass limit beyond which the formation of halos is inhibited. On the proton-rich side, it is believed that the Coulomb interaction
prevents the formation of halos beyond $Z\approx10$~\cite{jensen00}. However, this could be put into question as
non-trivial effects may come into play~\cite{liang07}. From a theoretical standpoint, halos in medium to heavy mass
nuclei can be studied through relativistic or non-relativistic Hartree-Fock-Bogoliubov
calculations~\cite{ring80a,bender03b} performed in the context of energy density functional (EDF) methods. On
the experimental side, the next generation of radioactive ion beam facilities
(FAIR at GSI, RIBF at RIKEN, REX-ISOLDE at CERN, SPIRAL2 at GANIL...) might be
able to assess the position of the neutron drip-line up to about $Z\approx26$~\cite{whitepapernscl}. Although this
would be an astonishing accomplishment, it will not allow the study of most of potential medium-mass halos. Still,
the (distant) future confrontation of theoretical results with experimental data will provide crucial information
that can be used to constrain theoretical models.

One difficulty resides in the absence of tools to characterize halo properties of finite many-fermion systems in a
quantitative way. Light nuclei constitute an exception considering that the quantification of halo properties in terms of
the dominance of a cluster configuration and of the probability of the weakly-bound clusters to extend beyond
the classical turning point is well acknowledged~\cite{fedorov93,jensen00,riisager00,jensen01}. Existing
definitions and tools applicable to systems constituted of tens of fermions are too qualitative, the associated observables are incomplete and have led to misinterpretations in nuclei~\cite{rotival07a}.

To improve on such a situation, a new quantitative and model-independent analysis method was proposed
in \refer\cite{rotival07a}. The method uses universal properties of the
internal one-body density to extract, in a model-independent fashion, the part of the density that can be
identified as a halo. Two criteria have been introduced to characterize halo systems in terms of (i) the
average number of fermions participating in the halo and (ii) the influence of the latter on the system
extension. The results deduced from EDF calculations of medium-mass nuclei have underlined the likely formation of a collective halo at the
neutron drip-line of chromium isotopes. The neutron density of those nuclei displays a spatially decorrelated region
built out of an admixture of $\ell=0$ and \mbox{$\ell=2$} overlap functions (orbitals). The significant
contribution of orbitals with orbital angular-momentum $\ell=2$ is at variance with the standard picture in light
nuclei~\cite{riisager92,mizu97}.

Given such an analysis method, important questions can now be addressed. The versatility of the method must be tested which we do in the first part of our study by applying it to many-fermion systems of different scales computed with various many-body techniques : atom-positron/ion-positronium complexes on the one hand and light/heavy nuclei on the other. In the second part of the article, we focus on the influence of (i) the scheme used to regularize/renormalize the ultra-violet divergence of the local
pairing functional, (ii) the angular-momentum cutoff in the single-particle basis, as well as (iii) the isoscalar effective
mass, (iv) saturation density of nuclear matter and (v) tensor terms~\cite{lesinski07a} characterizing the particle-hole part of the energy
functional. More generally, the different ways pairing correlations impact halo nuclei are
studied, e.g. the anti-halo effect~\cite{bennaceur99,bennaceur00} or the potential decorrelation of $\ell=0$ orbitals from the
pairing field~\cite{hamamoto03,hamamoto04,hamamoto05}.

The present paper is organized as follows. The analysis method proposed in Ref.~\cite{rotival07a} is briefly recalled in \sect\ref{sec:crit} whereas its versatility is highlighted in \sect\ref{sec:fb}. In \sect\ref{sec:check},
technical aspects of Skyrme-EDF calculations are provided and the dependence of halo predictions on some their ingredients is pointed out. Section \ref{sec:pairing} is devoted to discussing the effect of pairing correlations on
the formation of halos. Then, the sensibility of halo properties to the characteristics of the particle-hole part
of the functional is studied. A large scale study of potential halos among all spherical medium-mass nuclei is proposed in \sect\ref{sec:fullscale}. In conclusions, we discuss how halo systems could help constraining the
nuclear EDF and to which extent the present results can be related to data generated in the distant future by
radioactive ion beam facilities.

\section{Characterization of halo systems}
\label{sec:crit}

\subsection{Analysis method}
\label{sec:method}

 Anew quantitative analysis-method of halos in finite many-fermion systems was proposed in Ref.~\cite{rotival07a}. The starting point is a model-independent decomposition of the internal one-body density
\mbox{$\rho_{\mathrm{[1]}}(r)$} of spherical many-fermion systems in terms of spectroscopic amplitudes
\mbox{$\varphi_\nu(\vec{r}\,)$} and their radial components $\bar{\varphi}_{n_\nu \ell_\nu
j_\nu}(r)$~\cite{vanneck93,vanneck98b,shebeko06}
\begin{equation}
\rho_{\mathrm{[1]}}(\vec{r}\,)=\sum_{\nu}|\varphi_\nu(\vec{r}\,)|^2 =\sum_{n_\nu \ell_\nu
j_\nu}\frac{2j_\nu+1}{4\pi}|\bar{\varphi}_{n_\nu \ell_\nu j_\nu}(r)|^2 \, \, . \label{eq:decomp_density}
\end{equation}

The appearance of a halo in the $N$-body system, i.e. the part of the density that is spatially decorrelated from an a priori unknown core, was shown in Ref.~\cite{rotival07a} to be related to the existence of three typical energy scales in the excitation
spectrum of the \mbox{$(N\!-\!1)$-body} system. From a practical viewpoint, and using the internal one-body density as the only input, the method allows the extraction of the radius $r_0$ beyond which the halo, if it exists, is located.

With the radius $r_0$ at hand, two quantitative {\it halo factors} are introduced. First, the average number of fermions participating in the halo can be extracted through
\begin{equation}
\label{eq:def_nhalo} N_{\mathrm{halo}}\equiv4\pi\int_{r_0}^{+\infty}\!\!\rho(r)\,r^2\,dr\,.
\end{equation}
Second, the contribution of the halo region to the root mean square radius of the system is also extracted
\begin{eqnarray}
\delta R_{\mathrm{halo}}&\equiv&R_{\mathrm{r.m.s.,tot}}-R_{\mathrm{r.m.s.,inner}}\notag\\
&=&\!\!\!\!\sqrt{\frac{\int_{0}^{+\infty}\rho(r)r^4\,dr}
{\int_{0}^{+\infty}\rho(r)r^2\,dr}}-
\sqrt{\frac{\int_{0}^{r_0}\rho(r)r^4\,dr}
{\int_{0}^{r_0}\rho(r)r^2\,dr}}\,.
\label{eq:def_drhalo}
\end{eqnarray}

\subsection{Versatility of the method}
\label{sec:fb}

To illustrate the analysis method and its versatility, we now apply it to the results of many-body calculations performed for three
different systems:  light nuclei studied
through coupled-channels calculations~\cite{nunes96a,nunes96b}, medium-mass nuclei described through single-reference energy density functional calculations and atom-positron/ion-positronium complexes computed with the
fixed-core stochastic variational method~\cite{varga95,varga97,suzuki98, ryzhikh98b,mitroy01}.

\subsubsection{Light nuclei}

To check the consistency of the method in a situation where core and halo densities are explicitly computed, the
values of $N_{\mathrm{halo}}$ and $\delta R_{\mathrm{halo}}$ have been extracted from coupled-channels calculations of light nuclei. This will serve a basis for cases where the total density only is accessible.

Calculations are performed for a core+neutron system for which the internal dynamics of the core is taken into
account~\cite{nunes96a,nunes96b} and the total Hamiltonian reads as
\begin{equation}
{H}_{\mathrm{tot}}={H}_{\mathrm{core}}+{T}_{\mathrm{rel}}+{V}_{\mathrm{n-core}}\,.
\end{equation}

To provide adequate nuclear quadrupole couplings, a deformed Woods-Saxon potential in the core rest frame is
considered
\begin{eqnarray}
V_{\mathrm{WS}}(r,\theta)&=&\left[1+e^{\frac{r-R(\theta)}{a_{\mathrm{WS}}}}\right]^{-1} \, \, \, ,\\
R(\theta)&=&R_{\mathrm{WS}}\left(1+\beta \,Y_{2}^0(\theta)\right) \, \, \, ,
\end{eqnarray}
where $\beta$ is the core quadrupole deformation. The total wave-function is expanded in a basis of eigenstates of the total angular momentum using a separation of the core internal motion, with eigenstates associated to the energies $\epsilon_i$, from the neutron relative motion. The resulting coupled-channels equations for the loosely bound nucleon wave function $\Psi_i$ read
\begin{equation}
\left({T}_{\mathrm{rel}}+\epsilon_i-E\right) \, \Psi_i+\sum_{j}V_{ij}\, \Psi_j=0 \, \, \, ,
\end{equation}
and are solved in a Sturmian basis.

\begin{figure}[hptb]
\includegraphics[keepaspectratio, angle = -90, width = \columnwidth]{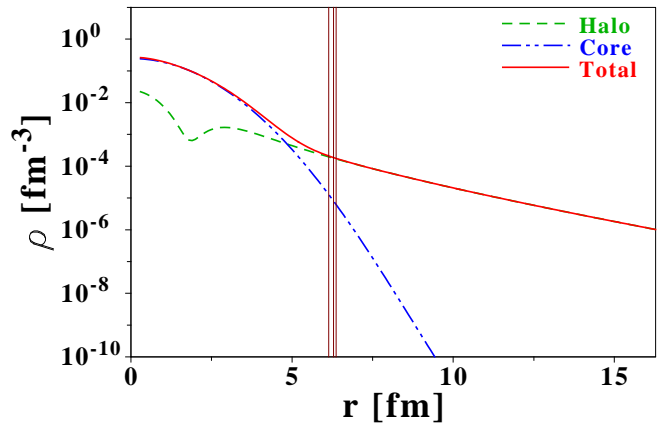}
\caption{(Color Online) Core, halo and total neutron density of \mbox{$^{11}$Be} obtained from coupled-channels
calculations~\cite{nunes96a,nunes96b}. The value of $r_0$ and the tolerance margin are indicted by vertical
lines.} \label{fig:Be}
\end{figure}

Calculations have been performed for two nuclei: the well-established one-neutron halo \mbox{$^{11}$Be}, whose
total density and core+halo decomposition are plotted in \figu\ref{fig:Be}, and the stable nucleus \mbox{$^{13}$C}
as a control case. The results are summarized in \tab\ref{tab:Be}, where the two criteria are evaluated for core
and total densities.
\begin{table}
 \setlength{\extrarowheight}{3pt}
\begin{tabular}{|c||cc|ccc|}
\hline
& \multicolumn{2}{c}{Core} &\multicolumn{3}{c|}{Total} \\
\hline
~~~~~ & $N_{\mathrm{halo}}$ & $\delta R_{\mathrm{halo}}$~[fm] & $N_{\mathrm{halo}}$ & $\delta R_{\mathrm{halo}}$~[fm]  & $R_{\mathrm{r.m.s.}}$~[fm] \\
\hline
$^{13}$C  & $0.000$ & $0.000$ & $0.66\times10^{-3}$ &  $0.74\times10^{-3}$ & $2.487$   \\
$^{11}$Be  & $0.000$ & $0.000$ & $0.270$ & $0.394$ & $2.908$   \\
\hline
\end{tabular}
\caption{\label{tab:Be} Values of $N_{\mathrm{halo}}$ and $\delta R_{\mathrm{halo}}$ for $^{13}$C (stable) and $^{11}$Be
(one-neutron halo). The criteria are applied to results obtained through coupled-channels
calculations~\cite{nunes96a,nunes96b}.}
\end{table}
For \mbox{$^{11}$Be}, $r_0$ is found to be compatible with the condition to have one order of magnitude difference
between core and halo densities at that radius~\cite{rotival07a}. The tail-to-core ratio is slightly different from ten, partly
because the core is represented by a gaussian profile with the wrong asymptotic. In any case, the ideal value of
$r_0$ still lies within the allowed theoretical error. This validates the method on a realistic system. The halo
parameter $N_{\mathrm{halo}}$ shows that around $0.3$ neutron reside in the decorrelated region in average. The reason why
one only finds a fraction of a neutron within the halo region is because the wave function of the "halo nucleon"
lies partly inside the volume of the core. The denomination of one-neutron halo is somewhat misleading from that
point of view. The influence $\delta R_{\mathrm{halo}}$ of the halo on the nuclear extension is large, of about $0.4$~fm
out of a total root-mean-square (r.m.s.) radius of $2.90$~fm. Note that such a value is very close to the value extracted
experimentally when going from $^{11}$Be to $^{10}$Be~\cite{tanihata88b}. Last but not least, $N_{\mathrm{halo}}$ and
$\delta R_{\mathrm{halo}}$ are found to be negligible for the two control cases considered, i.e. \mbox{$^{13}$C} and the
core density of \mbox{$^{11}$Be}. This illustrates the ability of the method to discriminate between halo and
non-halo systems.

\subsubsection{Medium-mass nuclei} \label{sec:hfb}

In Ref.~\cite{rotival07a}, the formation of neutron halos in Cr and Sn isotopes has been investigated through self-consistent EDF
calculations performed using the SLy4 Skyrme functional complemented with a density-dependent mixed-type pairing
functional (see \sect\ref{sec:skyrme_ph} and \sect\ref{sec:skyrme_pp}). The value of $r_0$ is determined
using as an input the one-body density obtained from a symmetry-breaking HFB state, which is assumed to map out the nuclear
internal density (see discussion in \refer\cite{rotival07a}).

The (perturbative) excitation spectra of the last three bound odd Cr isotopes \mbox{($Z=24$)} shown in Ref.~\cite{rotival07a}
display optimal energy scales as far as the formation of a halo is concerned. The values of
$N_{\mathrm{halo}}$ and $\delta R_{\mathrm{halo}}$ found for the last four bound even Cr isotopes, \mbox{$^{80}$Cr} being the
predicted neutron drip-line nucleus for the parametrization used, are listed in \tab\ref{tab:res_Cr}. Beyond the
\mbox{$N=50$} neutron shell closure, the steep increase of both $N_{\mathrm{halo}}$ and $\delta R_{\mathrm{halo}}$ is the signature
of a halo formation. A decorrelated region containing up to $\sim0.5$ neutron appears at the limit of stability.
Such a number of neutrons participating in the halo is small relatively to the size of the system but comparable
in absolute value to the number found in light halo nuclei, as recalled in \sect\ref{sec:fb}.

The contribution of the halo region to the nuclear extension reaches about $0.13$~fm in \mbox{$^{80}$Cr}. On the
one hand, such a value is significant in comparison with the total neutron r.m.s. radius of
those systems. On the other hand, it corresponds to about one-third of the value found for \mbox{$^{11}$Be}. It is likely that the formation of halos is hindered as the mass increases because of
the increased collectivity. We will come back to that in section~\ref{sec:fullscale}. Of course, one should not disregard the fact that single-reference EDF calculations miss certain long-range correlations which, once they are included, might slightly change the picture.

One interesting feature discussed in Ref.~\cite{rotival07a} is that the kink of the r.m.s. radius at \mbox{$N>50$}, which had been interpreted as a halo signature in previous works, is partly due to
a plain shell effect associated with the sudden drop of the two-neutron separation energy. As a matter of fact, the quantity $\delta R_{\mathrm{halo}}$ allows the disentanglement of shell and halo effects in the increase of the neutron r.m.s. radius across $N=50$. For isotopes further away from their drip-line, e.g. across $N=82$ in tin isotopes, the use of $\delta R_{\mathrm{halo}}$ demonstrates that the kink of the neutron r.m.s. radius is entirely produced by the shell effect, whereas its further increase beyond $N=82$ is related to the growth of the neutron skin, not to the appearance of a halo~\cite{rotival07a}. Generally speaking the quantity
$\delta R_{\mathrm{halo}}$ does not incorporate the contribution from the neutron skin and only characterizes the halo part of the
density profile, e.g. spatially decorrelated neutrons.

\begin{table}[htbp]
 \setlength{\extrarowheight}{3pt}
\begin{tabular}{|c||c|c|}
\hline
~~~~~~~~ & ~~~~~~$N_{\mathrm{halo}}$~~~~~~ & ~~$\delta R_{\mathrm{halo}}$~[fm]~~ \\
\hline
$^{74}$Cr & $0.000$ & $0.000$   \\
$^{76}$Cr & $0.057$ & $0.018$   \\
$^{78}$Cr & $0.194$ & $0.055$   \\
$^{80}$Cr & $0.472$ & $0.134$   \\
\hline
\end{tabular}
\caption{\label{tab:res_Cr} Values of $N_{\mathrm{halo}}$ and $\delta R_{\mathrm{halo}}$ for the last bound even-even Cr isotopes,
as predicted by the \{SLy4+REG-M\} functional. Results slightly differ from the ones shown in Paper I because of
the different codes used to perform the EDF calculations. The difference remains very small, of the order of the
third significative digit.}
\end{table}

A decomposition of the halo region in terms of canonical states shows that it contains equal contributions
from neutron $3s_{1/2}$ and $2d_{5/2}$ orbitals. The possibility for \mbox{$\ell\ge 2$} states to participate
in the halo was not rejected from the outset. Eventually, the loosely bound  $2d_{5/2}$ shell strongly contributes here because of its rather large occupation and intrinsic
degeneracy. The latter observations point towards the formation of a
collective halo in drip-line Cr isotopes, formed by an admixture of several overlap functions, and call for a
softening of the restrictions spelled out in light nuclei~\cite{jensen04}.

The energy scales necessary to the halo formation are not seen in the spectra of drip-line tin isotopes and the
weak kink observed in the neutron r.m.s. radius at \mbox{$N=82$} is nothing but a shell effect, as mentioned
above. In contrast with previous studies based on the Helm model
~\cite{helm56,rosen57,raphael70,friedrich82,mizutori00}, no collective halo is identified in tin isotopes, as the decorrelated region
is found to have almost no influence on the matter extension~\cite{rotival07a}.

\subsubsection{Atom-positron/ion-positronium complexes} \label{sec:apos}

In atomic physics, valence electrons of neutral atoms can be located at large distances from the core. Because of
the very long range of the Coulomb interaction, the penetration of the wave-function into the classically
forbidden part of the potential as the separation energy of the system becomes small cannot be interpreted as a
halo formation~\cite{jensen04}. However, a positron can be attached to a neutral atom by the polarization potential, which can be
parameterized as

\begin{eqnarray}
V_{\mathrm{pol,1}}(\vec{r}\,)&=&-\frac{\alpha_D\,g^2(r)}{2r^4}
\underset{r\rightarrow +\infty}{\longrightarrow}
- \frac{\alpha_D}{2r^4}\,,\\
g^2(r)&=&1-e^{-\frac{r^6}{\beta^6}}\,,
\end{eqnarray}
where $\alpha_D$ is the core polarization constant and $\beta$ a cutoff distance. In this case, the $r^{-4}$ decay
of the potential at large distances does not ensure that particles are able to tunnel through the potential
barrier. It was found that several atom-positron complexes can exist~\cite{ryzhikh97,ryzhikh98a,ryzhikh98b,
bromley02a,bromley02b,bromley02c,bromley02d,mitroy02}, and have been identified as having halo
characteristics~\cite{mitroy05}. To quantify such an observation, values of $N_{\mathrm{halo}}$ and $\delta R_{\mathrm{halo}}$ are
evaluated for such systems.

The Hamiltonian of the atom-positron system with $N_{\mathrm{val}}$ valence electrons reads with normalized units
(\mbox{$m_e=e=1$})~\cite{ryzhikh98a,ryzhikh98b}

\begin{eqnarray}
\hat{H}&=&\sum_{i=1}^{N_{\mathrm{val}}}\left(-\frac{1}{2}\vec{\nabla}_i^2+V_{\mathrm{dir}}(\vec{r}_i)+V_{\mathrm{ex}}(\vec{r}_i)
+V_{\mathrm{pol,1}}(\vec{r}_i)\right) \nonumber \\
&&~~~~+\sum_{\substack{i,j=1\\i<j}}^{N_{\mathrm{val}}}\left(\frac{1}{r_{ij}}-V_{\mathrm{pol,2}}(\vec{r}_i,\vec{r}_j)\right)\nonumber\\
&&~~~~~~~~-\frac{1}{2}\vec{\nabla}_0^2-V_{\mathrm{dir}}(\vec{r}_0)+V_{\mathrm{pol,1}}(\vec{r}_0)\nonumber\\
&&~~~~~~~~~~~~-\sum_{i=1}^{N_{\mathrm{val}}}\left(\frac{1}{|\vec{r}_i-\vec{r}_0|}-V_{\mathrm{pol,2}}(\vec{r}_i,\vec{r}_0)\right) \, , \,
\end{eqnarray}
where $\vec{r}_0$ is the positron position vector, \mbox{$\vec{r}_{ij}=\vec{r}_i-\vec{r}_j$} is the relative position of
two valence electrons, whereas the direct $V_{\mathrm{dir}}$ and exchange $V_{\mathrm{ex}}$ potentials between valence electrons and
the core are computed exactly in the Hartree-Fock approximation.
 The two-body polarization potential is defined as
\begin{equation}
V_{\mathrm{pol,2}}(\vec{r}_i,\vec{r}_j)=\frac{\alpha_d}{r_i^3\,r_j^3}(\vec{r}_i.\vec{r}_j)\,g(r_i)\,g(r_j)\,.
\end{equation}
\begin{table*}
 \setlength{\extrarowheight}{3pt}
\begin{tabular}{|cc||cc||cc||cc||ccc|}
\hline
~~Atom~~ & ~~Asympt.~~ & ~~$N_{e^-}$~~ &  ~~$N_{\mathrm{halo}}$~~ & ~~$R_{\mathrm{r.m.s.}}$~[$a_0$]~~ &
~~$\delta R_{\mathrm{halo}}$~[$a_0$]~~& ~~$P_{e^+}$~[\%]~~ & ~~$P_{e^-}$~[\%]~~ & $E_{\mathrm{gs}}$~[at. units]~~ & ~~$\epsilon$~[at. units]~~ & \refer \\
\hline
Be  & $e^+$+Be  & $4$  & $0.624$ &  $5.661$  & $3.194$  & $98.1$ & $01.9$ & $-1.0151$ & $0.0032$ & \cite{mitroy05} \\
Mg  & $e^+$+Mg  & $12$ & $0.669$ &  $2.298$  & $0.826$  & $80.3$ & $19.7$ & $-0.8477$ & $0.0156$ & \cite{mitroy01} \\
Cu  & $e^+$+Cu  & $29$ & $0.754$ &  $1.777$  & $0.975$  & $88.6$ & $11.4$ & $-0.2891$ & $0.0051$ & \cite{bromley02d} \\
\hline
He  & Ps+He$^+$ & $2$  & $1.982$ &  $15.472$ & $14.568$ & $50.3$ & $49.7$ & $-2.2506$ & $0.0006$ & \cite{mitroy05b} \\
Li  & Ps+Li$^+$ & $3$  & $1.972$ &  $7.781$  & $7.088$  & $50.8$ & $49.2$ & $-7.5324$ & $0.0024$ & \cite{mitroy04} \\
\hline
\end{tabular}
\caption{\label{tab:cplx} Results of the halo analysis for various atom-positron systems evaluated with the fixed-core
stochastic variational method. The columns provide: neutral atom symbol, asymptotic form, total number of
electrons $N_{e^-}$, halo factors $N_{\mathrm{halo}}$ and $\delta R_{\mathrm{halo}}$, total matter r.m.s. radius $R_{\mathrm{r.m.s.}}$, relative
proportions $P_{e^+}$ and $P_{e^-}$ of electrons and positrons in the halo region, ground state energy $E_{\mathrm{gs}}$ of
the \mbox{$e^+$+A} complex, and binding energy $\epsilon$ with respect to the corresponding dissociation
threshold. All length units are normalized to the Bohr radius $a_0$, and energy values are in atomic units
\mbox{($1$~at. unit\,$=27.21162$~eV)}.}
\end{table*}
\begin{figure}[hptb]
\includegraphics[keepaspectratio, angle = -90, width = \columnwidth]{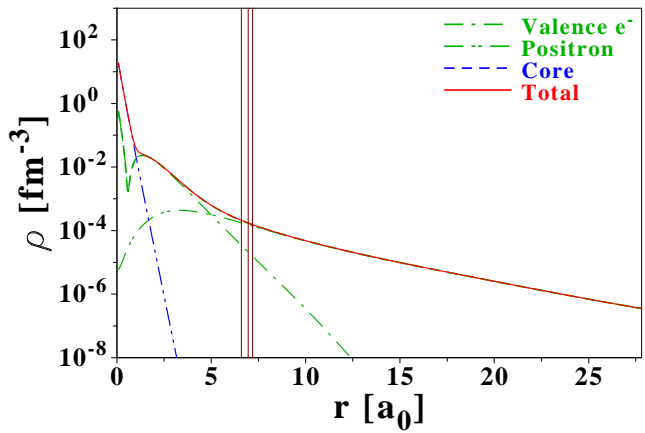}
\caption{(Color Online) Valence electron, positron, core and total density for the \mbox{$e^++$Be} complex evaluated with
the fixed-core stochastic variational method. All length units are normalized to the Bohr
radius $a_0$. The value of $r_0$ and the tolerance margin are indicated by vertical lines.} \label{fig:Be_posit}
\end{figure}
\begin{figure}[hptb]
\includegraphics[keepaspectratio, angle = -90, width = \columnwidth]{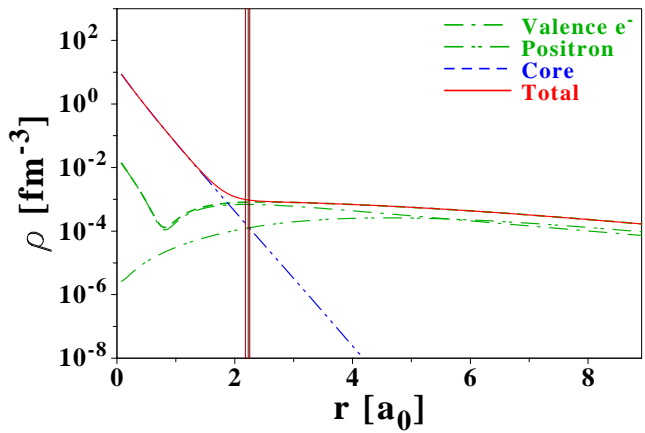}
\caption{(Color Online) Same as in \figu\ref{fig:Be_posit} for the \mbox{$e^+$+Li} system.} \label{fig:Li_posit}
\end{figure}

When the system \mbox{$e^{+}$+A} is bound, its asymptotic behavior can correspond to (i) a neutral core $A$ plus a
positron $e^+$, or (ii) a charged core $A^+$ and a neutral positronium complex Ps, depending on the relative
binding energies of those configurations. Calculations are performed with the fixed-core stochastic variational
Method (FCSVM)~\cite{varga95,varga97,suzuki98, ryzhikh98b,mitroy01}, in a basis of explicitly correlated gaussians
for the individual wave-functions. The basis is taken large enough to correctly reproduce the asymptotic behavior
of the \mbox{$e^+$+A} or \mbox{Ps+A$^+$} systems. The results of such calculations for \mbox{$e^+$+Be},
corresponding to the \mbox{$e^+$Be} complex, are presented in \figu\ref{fig:Be_posit}, where the separation at
large distances between a weakly bound positron and a core composed of the electrons is visible. As a result, a
positron extended tail appears. In \figu\ref{fig:Li_posit} are also displayed the results for the \mbox{$e^+$+Li}
system, which corresponds to a \mbox{PsLi$^+$} complex. Indeed, one observes that the density tail is
composed of almost identical $e^+$ and $e^-$ components.

The results of the analysis, performed for several atom-positron complexes, are presented in \tab\ref{tab:cplx}.
The separation energy $\epsilon$ in the appropriate channel (\mbox{$e^+$+A} or \mbox{Ps+A$^+$}) is small compared
to the ground state energy $E_{\mathrm{gs}}$ of the complex. The situation regarding the energy scales at play is very
favorable as far as the formation of halos is concerned. It is also possible to evaluate the composition of the
halo region in terms of the proportion of electrons $P_{e^-}$ and of positrons $P_{e^+}$.

The values of $N_{\mathrm{halo}}$ and $\delta R_{\mathrm{halo}}$ demonstrate the existence of halos in \mbox{$e^+$+Be},
\mbox{$e^+$+Mg} and \mbox{$e^+$+Cu} which strongly affect the system extensions. For example, the spatially
decorrelated part of the density accounts for about half of the total r.m.s. radius in \mbox{$e^+$+Be}, although
it contains only $\sim0.7$ particle in average. In those cases, a positron halo is predicted as the halo region is
almost exclusively built from the positron wave-function (\mbox{$P_{e^+}\gg P_{e^-}$}). For
\mbox{$e^+$+Li} and \mbox{$e^+$+He}, extremely large values for $N_{\mathrm{halo}}$ and $\delta R_{\mathrm{halo}}$ are extracted, demonstrating that one is dealing in these cases with gigantic ion-positronium halos (\mbox{$P_{e^+}\approx P_{e^-}$}).

Considering the values of $N_{\mathrm{halo}}$ and $\delta R_{\mathrm{halo}}$, one realizes that atom-positron and ion-positronium
complexes display more extreme halo structures than nuclei. This is of course due to the nature of the
interactions at play in such systems.

\subsection{Universality of the phenomenon}
\label{sec:univ}

As demonstrated, the method developed in Ref.~\cite{rotival07a} can be applied successfully to finite many-fermion systems of very different scales. This relates to the fact that the method relies on a model-independent analysis of the internal one-body density
\mbox{$\rho_{\mathrm{[1]}}(\vec{r}\,)$}. In all cases, a fraction of the constituents may extend far out from the core and influence
strongly the size of the system.

\begin{figure}[hptb]
\includegraphics[keepaspectratio, angle = -90, width=\columnwidth]{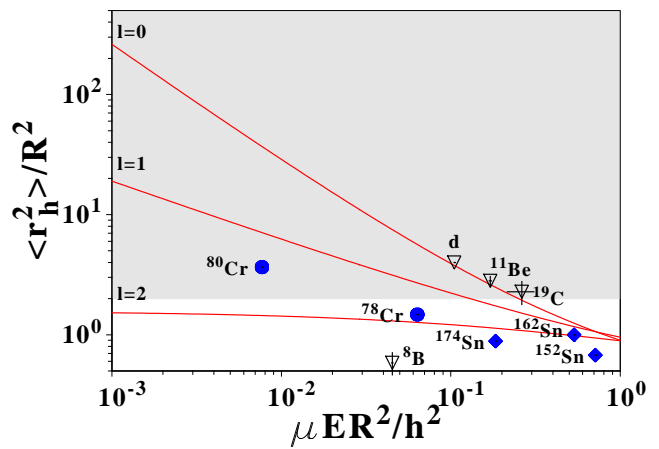}
\caption{(Color Online) Universality of halos features. The dimensionless extension of the halo is plotted against the
dimensionless separation energy for EDF calculations of medium-mass nuclei (filled symbols) and experimental results for few-body systems
(open symbols). Generic scaling relationships obtained from a finite spherical well are given in solid lines,
while the gray-shaded area corresponds to the halo are, as defined by the commonly-used criterion for few-body halo nuclei \mbox{$\langle
r_{h}^2\rangle/R^2>2$}. \label{fig:scaling}}
\end{figure}

It happens that halo systems display scaling properties which do not depend on their dimension and constituency.
This can be characterized by the extension of the halo wave-function \mbox{$\langle r^2_{h}\rangle$}
as a function of the separation energy $E$, and those quantities can be made dimensionless using as a scale
the classical turning point $R_q$ for the interaction potential of interest and the reduced mass of the
systems $\mu$~\cite{fedorov93,fedorov94,riisager00,jensen04}. The generic asymptotic scaling laws of
the two-body system are extracted using a finite spherical well, and depend on the angular momentum of the
weakly-bound overlap function~\cite{fedorov93}, as seen in \figu\ref{fig:scaling}. Results for light nuclei obey
rather well such universal scaling laws. For few-body systems, it is commonly admitted that halos appear when
\mbox{$\langle r_{h}^2\rangle/R_q^2>2$}, which corresponds to a probability greater than $50\,\%$ for the
weakly-bound nucleon to be in the forbidden region~\cite{jensen04}.

The results obtained for medium-mass nuclei can also be displayed in \figu\ref{fig:scaling} and compared with the
generic scaling laws. However, the dimensionless quantities have to be redefined. For medium-mass systems the halo
r.m.s. radius is evaluated, by analogy with \refer\cite{riisager00}, through
\begin{equation}
\langle r^2_h \rangle=\frac{N}{\mu}\,\langle r^2_{\mathrm{tot}}\rangle-\frac{N-N_{\mathrm{halo}}}{\mu}\,\langle r^2_{\mathrm{core}} \rangle\,,
\end{equation}
where \mbox{$\langle r^2_{\mathrm{tot}}\rangle$} is the total neutron r.m.s. radius, while the core r.m.s. radius is
approximated by $R_{\mathrm{r.m.s.,inner}}$ (see \eq(\ref{eq:def_drhalo})). The reduced mass is taken as the effective
isoscalar nucleon mass $m^\ast$, while the classical turning point $R_{q}$ of the central part $U_{q}(r)$ of the
one-body potential is evaluated, by analogy with the finite-well potential, as~\cite{fedorov93}
\begin{equation}
\frac{\displaystyle\int \,dr\,r^3\,U_{q}(r)}{\displaystyle\int\,dr\,r\,U_{q}(r)} = \frac{R^2_{q}}{2}\,.
\end{equation}
where $q=n$ as we are interested in neutron halos.

In fig.~\ref{fig:scaling}, the last bound Cr isotopes are located in-between the \mbox{$\ell=0$} and \mbox{$\ell=2$} scaling curves, with
$^{78}$Cr being closer to the \mbox{$\ell=2$} curve than $^{80}$Cr. This is consistent with the admixture of
orbitals that builds the corresponding halos, as discussed in \sect\ref{sec:hfb}~\cite{rotival07a}. The neutron
 density of most medium-mass halo nuclei does not extend as much as those of few-body systems such as \mbox{$^{11}$Be}.
 Still, \mbox{$^{78}$Cr} and \mbox{$^{80}$Cr} display few-body-like halo properties and the ratio
 \mbox{$\langle r_{h}^2\rangle/R_n^2$} does exceed $2$ for \mbox{$^{80}$Cr}. On the contrary, the
 extension of neutron-rich tin isotopes is not significant enough in regard with their separation energy to be characterized
 as halo systems. This is consistent with the findings of Paper I.

It would be of interest to place atom-positron complexes in \figu\ref{fig:scaling}. However, the many-body method
used to compute their properties does not allow an easy extraction of the corresponding classical turning point.
As a result, the corresponding results do not appear in \figu\ref{fig:scaling}.

\section{Technical aspects}
\label{sec:check}

The objectives of the remaining part of the present study are (i) to predict which spherical medium-mass nuclei might display halo features,
(ii) to check the dependence of the results on several ingredients of the numerical implementation, (iii) to probe the
sensitivity of the predictions to the characteristics of the many-body treatment and (iv) to study the specific
impact of pairing correlations on halo nuclei.

In the present section, and because halos are extreme systems whose asymptotic must be properly accounted for, we study the dependence of our predictions on certain technical features of the many-body calculation. To do so, ingredients of single-reference EDF calculations are briefly recalled at first.

\subsection{HFB equations}
\label{sec:HFB}

Spherical symmetry is assumed throughout the rest of the paper and spin/isospin indices are sometimes omitted for simplicity. Calculations are performed using a code that takes advantage of the so called ``two-basis method'' to solve
the HFB equations~\cite{gall94}. Thanks to the spherical symmetry, the HFB equation are solved for each $(l,j)$ block separately in the
basis $\{\phi_{nlj}\}$ that diagonalizes the single-particle field $h$

\begin{equation}
\sum_{k}\left[\begin{matrix}
\left( h^{lj}_{ik} - \lambda\right) \delta_{ik}  & \tilde{h}^{lj}_{ik} \\[2mm]
\tilde{h}^{lj}_{ik} &- \left( h^{lj}_{ik} - \lambda\right) \delta_{ik}
\end{matrix}\right]
\left[\begin{matrix}
U^{lj}_{kn} \\[2mm]
V^{lj}_{kn}
\end{matrix}\right]
=E_{nlj} \left[\begin{matrix}
U^{lj}_{in} \\[2mm]
V^{lj}_{in}
\end{matrix}\right] \, ,
\label{eq:hfbb}
\end{equation}
where $U^{lj}_{kn}$ and $V^{lj}_{kn}$ are the expansion coefficients of the upper and lower parts of the HFB
spinor $\Psi_{nlj}$ on the basis \mbox{$\{|\phi_{klj}\rangle\}$}, whereas $\tilde{h}$ denotes the pairing field. The two-basis method
authorizes to perform calculations in very large boxes. The nucleus is put in a spherical box such that wave functions are computed up to a radial distance $R_{\mathrm{box}}$, with vanishing boundary conditions (Dirichlet) imposed. The value of $R_{\mathrm{box}}$ has to be
chosen large enough as to ensure convergence of the calculations (see \sect\ref{sec:cvg_tests}). The differential equation to find the $\{\phi_{nlj}\}$
is solved on a discrete mesh of step size $h=0.25$~fm using the Numerov algorithm~\cite{dahlquist74,bennaceur05a}.

In \eq(\ref{eq:hfbb}) the upper bound of the sum over the index $k$ is not specified. In an actual
calculation, the sum is truncated by keeping states $\{\phi_{nlj}\}$ up to a certain maximum energy
$E_{\mathrm{max}}$. Its actual value ranges from several MeV up to hundreds of MeVs depending on the method used to
tackle the ultra-violet divergence of the local pairing functional (see section~\ref{sec:skyrme_ph}). In addition to the energy cut, a truncation $j^q_{cut}$ on the number of
partial waves kept in the basis $\{\phi_{nlj}\}$ is implemented. In principle, all wave functions below
$E_{\mathrm{max}}$ should be kept, but this makes the computation time rather long whereas the wave functions with
very high angular momenta will not contribute to the nuclear density. Nonetheless, such a truncation must not be
too drastic, in particular for loosely bound nuclei. Checks of convergence with
respect to $R_{\mathrm{box}}$, $E_{\mathrm{max}}$ and $j^q_{\mathrm{cut}}$ of observables of interest in halo nuclei are discussed in \sect\ref{sec:cvg_tests}.

\subsection{Energy density functional}
\label{sec:EDF}

\subsubsection{Particle-hole channel}
\label{sec:skyrme_ph}

The Skyrme part of the EDF takes, in the case of spherical nuclei, the standard form~\cite{bender03b}

\begin{align}
{\cal E}^{\rho\rho}=\frac{\hbar^2}{2m}\tau_0 +\sum_{T=0,1}&C_T^\rho\rho_T^2 +C_T^{\Delta\rho}\rho_T\,\Delta\rho_T
+C_T^{\tau}\rho_T\tau_T \nonumber\\
&+\chi \,C_T^{J}{\vec{J}}_T^{\, 2} +C_T^{\Delta J}\rho_T{\vec{\nabla}}\cdot{\vec{J}}_T\,, \label{eq:Vph}
\end{align}
where $\rho(\vec{r}\,)$, $\tau(\vec{r}\,)$ and $\tensor{J}(\vec{r}\,)$ denote the normal, kinetic and spin-orbit
densities, respectively. The parameter $\chi$ is equal to zero or one, depending on whether
tensor terms are included or not in the functional, while the index $T$ labels isoscalar
\mbox{($T=0$)} and isovector \mbox{($T=1$)} densities. For protons, the
particle-hole part of the EDF is complemented with a Coulomb term whose exchange part is treated within the
Slater approximation~\cite{bender03b}.

To study the effect of specific features of the particle-hole functional on the formation of halos, a set of Skyrme functionals characterized by
different properties is used in the present study: (i) SLy4 stands as a reference point, (ii) SIII displays a
different density dependence which leads to a too high infinite matter incompressibility $K_{\infty}$, (iii) T6 has
an isoscalar effective nucleon mass \mbox{$(m^\ast/m)_s=1$}, providing a denser single-particle spectrum, (iv) SKa
has a low isoscalar effective mass and a different density dependence (density-dependent term with an
exponent of \mbox{${1}/{3}$} instead of \mbox{${1}/{6}$}), (v) the functional ``$m^\ast 1$'' has been specifically
adjusted for the present work with the same procedure as for SLy4 but with the constraint \mbox{$(m^\ast/m)_s=1$},
(vi) the parameterizations ``$\rho_{\mathrm{sat}}^{1/2/3}$'' have also been adjusted specifically, with different nuclear
matter saturation densities $\rho_{\mathrm{sat}}$, (vii) T21 to T26 incorporate tensor terms that differ by their
neutron-neutron couplings~\cite{lesinski07a}. The parameterizations ``$m^\ast 1$'' and ``$\rho_{\mathrm{sat}}^{1/2/3}$''
have been adjusted using the procedure of \refer\cite{lesinski07a} which amounts to reproducing (i) the binding
energies and charge radii of \mbox{$^{40,48}$Ca}, \mbox{$^{56}$Ni}, \mbox{$^{90}$Zr}, \mbox{$^{132}$Sn} and
\mbox{$^{208}$Pb}, (ii) the binding energy of \mbox{$^{100}$Sn}, and (iii) the equation of state of pure neutron
matter~\cite{wiringa95} and other standard properties of symmetric nuclear matter as well as the
Thomas-Reiche-Kuhn enhancement factor of the isovector giant dipole resonance.

Infinite matter properties of all used parameterizations are summarized in \tab\ref{tab:skyrme_prop}. The
isovector effective mass (related to the Thomas-Reiche-Kuhn enhancement factor $\kappa_v$) is significantly
different for these parameterizations but its effect on static properties of nuclei is rather
small~\cite{lesinski06a}.

\begin{table}
 \setlength{\extrarowheight}{3pt}
\begin{tabular}{|c||ccccc|c|}
\hline
 & $\rho_{\mathrm{sat}}$ & $K_{\infty}$ & $(m^\ast/m)_s$ & $\kappa_v$ & $E/A$ & \refer \\
\hline
\hline
SLy4             & \,$0.160$\, & \,$229.9$\, & $0.70$ & 0.25 & \,$-15.97$\, & \cite{chabanat97,chabanat98}\\
SIII             & $0.145$ & $355.4$ & $0.76$ & 0.53 & $-15.85$   & \cite{beiner75a}   \\
$m^\ast 1$           & $0.162$ & $230.0$ & $1.00$ & 0.25 & $-16.07$   & \cite{lesinskipriv}   \\
$\rho_{\mathrm{sat}}^1$   & $0.145$ & $230.0$ & $0.70$ & 0.25 & $-15.69$   & \cite{lesinskipriv}   \\
$\rho_{\mathrm{sat}}^2$   & $0.160$ & $230.0$ & $0.70$ & 0.25 & $-15.99$   & \cite{lesinskipriv}   \\
$\rho_{\mathrm{sat}}^3$   & $0.175$ & $230.0$ & $0.70$ & 0.25 & $-16.22$   & \cite{lesinskipriv}   \\
T6               & $0.161$ & $235.6$ & $1.00$ & 0.00 & $-15.93$ & \cite{tondeur84}   \\
SKa              & $0.155$ & $263.1$ & $0.61$ & 0.94 & $-15.99$ & \cite{kohler76}   \\
T21-T26          & $0.161$ & $230.0$ & $0.70$ & 0.25 & $-16.00$ & \cite{lesinski07a}   \\
\hline
\end{tabular}
\caption{\label{tab:skyrme_prop} Infinite matter properties of Skyrme functionals used in the present study:
saturation density $\rho_{\mathrm{sat}}$~[fm$^{-3}$], bulk compressibility $K_{\infty}$~[MeV], isoscalar effective mass
\mbox{$(m^\ast/m)_s$}, Thomas-Reiche-Kuhn enhancement factor $\kappa_v$ and energy per particle at saturation
$E/A$~[MeV].}
\end{table}

\subsubsection{Particle-particle channel}
\label{sec:skyrme_pp}

The local pairing functional used

\begin{equation}
{\cal E}^{\kappa\kappa} = \frac{V_0}{4} \, \left[ 1-\eta\left(\frac{\rho_0}{\rho_{\mathrm{sat}}} \right)^{\alpha}
\right] \sum_{q}
  \left|\trho^{q} (\vec{r}\,)\right|^{2}\, \, , \label{eq:Vpp}
\end{equation}
derives from a Density-Dependent Delta Interaction
(DDDI) with the same strength V0 for neutron-neutron
and proton-proton pairing. In addition, two parameters $\eta$ and $\alpha$ control the spatial dependence of the effective
coupling constant. With $\rho_{\mathrm{sat}}$ designating the saturation density of infinite nuclear matter, a zero value of $\eta$ corresponds to a pairing strength that is uniform over the nuclear volume (``volume
pairing'') while \mbox{$\eta=1$} corresponds to pairing strength which is stronger in the vicinity of the nuclear
surface (``surface pairing''). A value $\eta=1/2$ corresponds to an intermediate situation (``mixed-type
pairing''). The parameter $\alpha$ is usually set to one. Values
\mbox{$\alpha<1$} correspond to stronger pairing correlations at low density. In the present work, we are
interested in varying such empirical parameters over a large interval of values to quantify how much the
characteristics of the pairing functional impact halo systems. Note finally that, all along this work, the
strength $V_0$ is chosen so that the neutron spectral gap \mbox{$\langle\Delta_\kappa^n\rangle$}~\cite{doba96}
equals $1.250$~MeV for \mbox{$^{120}$Sn}. Such a value of $V_0$ provides
reasonable gaps in Ca, Sn and Pb regions.

To compensate for the ultra-violet divergence of the pairing density generated by the local pairing functional, a common procedure consists of regularizing all integrals at play through the use of a cutoff~\cite{doba84a}, e.g. on quasiparticle energies \mbox{$E_i<E_{\mathrm{cut}}$}. Pairing functionals using such a regularization scheme are referred to as
``REG'' in the following. In particular REG-S, REG-M and REG-V denote regularized surface-, mixed- and
volume-type pairing functionals, respectively. If the parameter $\eta$ differs from 0, 1/2 and 1 or if $\alpha$
differs from 1 the functional is generically noted as REG-X.

\begin{figure}[hptb]
\includegraphics[keepaspectratio, ,angle=-90, width = \columnwidth]{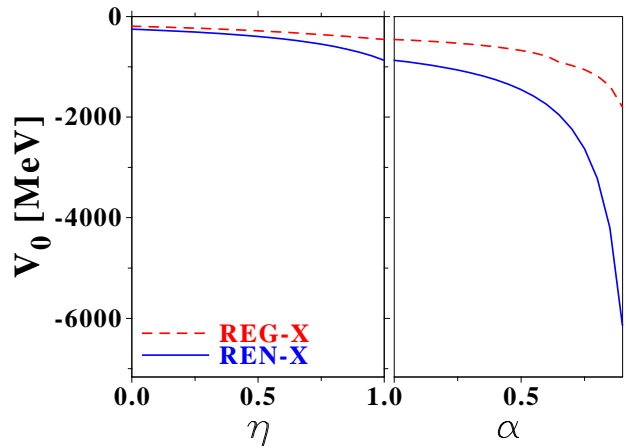}
\caption{(Color Online) Pairing strength $V_0$ as a function of $\eta$ and $\alpha$ for regularized REG-X and in-medium
renormalized REN-X pairing functionals.\label{fig:gap_adjust}}
\end{figure}

Using such a regularization method, the pairing strength is adjusted for each cutoff $E_{\mathrm{cut}}$, the latter being
eventually taken large enough for observables to be insensitive to its precise value. A widely used value is
\mbox{$E_{\mathrm{cut}}=60$~MeV}~\cite{doba84a,doba96,borycki06}. As the density dependence of the pairing functional is made more surface-peaked, the pairing strength
increases, for a fixed value of $E_{\mathrm{cut}}$, as is exemplified in \figu\ref{fig:gap_adjust}. This is a
consequence of the fitting procedure that uses a single nucleus to adjust the overall strength. Indeed, if the
pairing strength is peaked at the nuclear surface, individual gaps decrease, especially for well-bound orbitals
residing in the nuclear interior. To compensate for this effect and maintain the same value of
\mbox{$\langle\Delta^n_\kappa\rangle$} in one given nucleus, the overall pairing strength has to be increased.

As an alternative to the sharp cut-off regularization, one can identify and throw away the diverging part of the pairing density through the use of an auxiliary
quantity that diverges in the same way but from which the divergence can be subtracted analytically. We follow here the procedure introduced in
\refers\cite{bulgac99,bulgac01,bulgac,bulgac1,bulgac3}. Using such a renormalization procedure, results become independent of $E_{\mathrm{cut}}$ reaches about a few tens of MeVs~\cite{bulgac1}. In the following, calculations of finite nuclei are
performed using a conservative value of \mbox{$E_{\mathrm{cut}}=200$}~MeV (see
\sect\ref{sec:cvg_emax}). Pairing functionals combined with the renormalization scheme are referred
to as ``REN'' in the following. In particular REN-S, REN-M and REN-V denote renormalized surface-, mixed-
and volume-type functionals, respectively. If the parameter $\eta$ differs from 0, 1/2 and 1 or if $\alpha$
differs from 1 the functional is generically noted as REN-X.

\subsection{Regularization versus renormalization scheme}
\label{sec:pathos_finite}

It was observed in Ref.~\cite{doba01b} that combining the sharp cut-off regularization method with an extreme surface pairing functional $(\alpha<1)$ might lead to unrealistic matter and pairing densities. As our goal is to study the influence of the pairing functional attributes on halo properties, it is of importance to characterize such a feature in more detail. As a matter of fact, it can be shown that extreme surface REG-X functionals lead to a spurious Bardeen-Cooper-Schrieffer (BCS) Bose-Einstein-Condensation (BEC) phase transition of infinite matter. On the contrary such a nonphysical transition, associated with the use of a pairing functional that (wrongly) predicts a bound di-neutron system in the zero-density limit~\cite{baldo95}, does not occur with extreme surface REN-X functionals~\cite{rotivalxx}.

In the present section, we briefly explain that such a spurious feature is in one-to-one correspondence with producing unphysical matter and pairing densities through HFB calculations of finite nuclei. To do so, properties of the last bound Cr isotopes are calculated using different pairing
functionals (i) ($\eta\leq1,\alpha=1$), which covers from volume to standard surface pairing functionals, and (ii)
($\eta=1,\alpha\leq1$), which correspond to extreme surface pairing functionals.

\begin{figure}[hptb]
\includegraphics[keepaspectratio, angle = -90, width = \columnwidth]%
{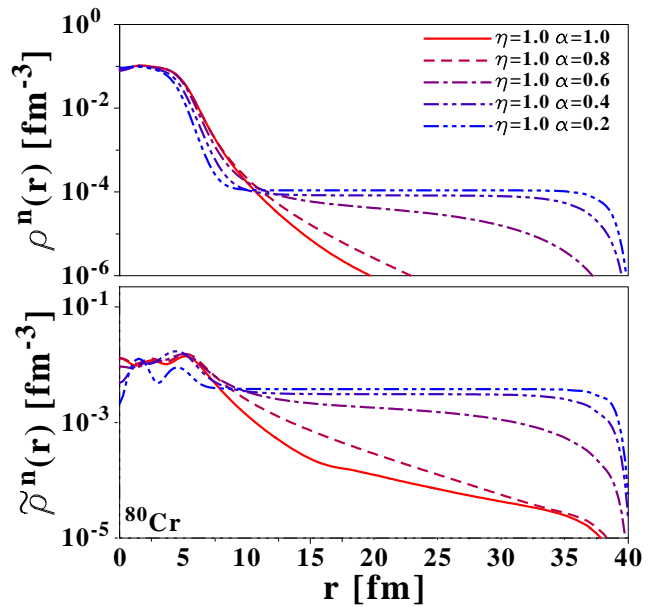} \caption{(Color Online) Neutron normal \mbox{$\rho^{n}(r)$} and pairing \mbox{$\trho^{n}(r)$} densities
plotted in logarithmic scale for \mbox{$^{80}$Cr} and several regularized REG-X pairing functionals. The
calculations are done in a box of $40$~fm and using all partial waves up to
$j^n_{\mathrm{cut}}=j^p_{\mathrm{cut}}=J_{\mathrm{max}}=151/2$.\label{fig:Cr80_dft_dens_1}}
\end{figure}

When the regularization scheme is used, one first observes that the position of the drip-line seems to change
drastically for extreme surface pairing functionals. For instance, \mbox{$^{84}$Cr} is predicted to be bound
against neutron emission $(\lambda^{q}<0)$~\cite{doba84a} for \mbox{$\alpha=0.5$}. However, the situation is more
subtle than it looks at first. As exemplified by \figu\ref{fig:Cr80_dft_dens_1} for \mbox{$^{80}$Cr}, nuclei are
in fact no longer bound as a gas of low-density superfluid neutron matter develops as \mbox{$\alpha$} decreases.
The normal and pairing (neutron) densities grow at long distances and become uniform beyond a radius $r\approx9$~fm
for $\alpha \lesssim 0.5$ which is precisely the critical value for which infinite nuclear matter undergoes an unphysical BCS-BEC transition~\cite{rotivalxx}. It becomes in fact energetically favorable for the nucleus to drip bound di-neutrons and create a superfluid gas.

Two remarks are at play at this point. First, it is important to use enough partial waves in the single-particle basis to describe the gas properly. Otherwise, one may not resolve it and conclude that the nucleus displays unusually extended normal and pairing densities~\cite{doba01b}. Here, calculations are performed including all partial waves up to $j^n_{\mathrm{cut}}=j^p_{\mathrm{cut}}=J_{\mathrm{max}}=151/2$. This is of course an extreme situation and less partial waves, but not too few, have to be used to describe genuine halos as is explained in the next section. Second, the appearance of a uniform gas of bound di-neutrons is driven by the neutron-neutron pairing interaction and not
by the proximity of the continuum. As a matter of fact, the fictitious nucleus whose densities are displayed on Fig.~\ref{fig:Cr80_dft_dens_1}, and which is held by the box, is still
bound against single-nucleon emission ($\lambda^q < 0$). If we were to increase $R_{\mathrm{box}}$, the system would turn into a gas of di-neutrons with binding energy $2 \lambda^q$~\cite{baldo95}.

\begin{figure}[hptb]
\includegraphics[keepaspectratio, angle = -90, width = \columnwidth]%
{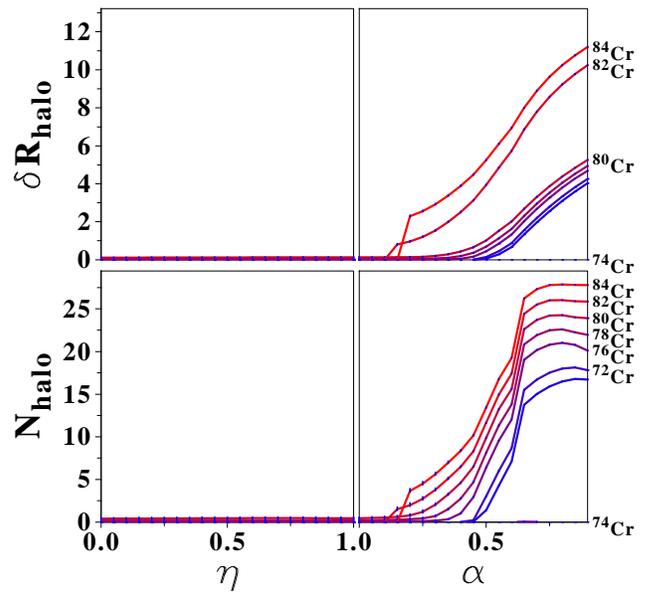} \caption{(Color Online) $N_{\mathrm{halo}}$ and $\delta R_{\mathrm{halo}}$ parameters for Cr isotopes and
different regularized REG-X pairing functionals.\label{fig:SV_DFTX}}
\end{figure}

Finally, one concludes that using a functional which wrongly predicts the existence of a bound di-nucleon state in
the $^{1}S_{0}$ channel in the vacuum translates into the creation of a spurious low-density di-neutron gas
surrounding finite nuclei. Such an observation is of importance regarding the analysis of halo systems as it
signals that the use of strongly surface-peaked pairing functionals, combined with the regularized scheme, might
lead to un-physical predictions. The danger resides in particular in the use of pairing functionals which are not
obviously un-physical, i.e. for which the di-neutron gas is not yet fully developed\footnoteb{Most probably because
of inappropriate numerical parameters.}. In such cases, the calculation will lead to wrongly predict the existence
of gigantic halos, e.g. see in \figu\ref{fig:SV_DFTX} the unreasonable values of $N_{\mathrm{halo}}$ and $\delta R_{\mathrm{halo}}$
predicted for the last bound Cr isotopes by REG-X functionals with $\alpha < 1$.

From that point of view, the renormalization scheme is safer as it prevents the problem from happening. Indeed, normal and pair neutron densities remain localized and evolve very little as \mbox{$\alpha<1$} decreases~\cite{rotivalxx}. On the other hand, it is crucial to point out that no problem occurs with the regularization scheme either as long as the low-density part of the pairing functional is physically constrained, e.g. $(\eta,\alpha)$ are adjusted as to reproduce pairing gaps calculated in infinite nuclear matter through ab-initio calculations~\cite{garrido2,matsuo06}. Eventually, the standard fitting strategy used here is the real source of the potential problem rather than the regularization method itself.

\subsection{Convergence checks} \label{sec:cvg_tests}

Several basis
truncations are introduced under the form of (i) a box of finite radius $R_{\mathrm{box}}$, (ii) an angular-momentum
cutoff $j_{\mathrm{cut}}^q$ for each isospin and (iii) a continuum energy cutoff $E_{\mathrm{max}}$, to accelerate the convergence of the
calculations. Such truncations are physically motivated as (i) the nucleus is localized in space (ii) high-lying
unoccupied states are not expected to contribute to ground state properties. However, the values of the truncation
parameters have to be carefully chosen not to exclude meaningful physics. As a result, the convergence of
observables of interest has to be checked, which we do now. All tests are performed for the chromium isotopes using the \{SLy4+REN-M\} functional.

\subsubsection{Box radius} \label{sec:cvg_rbox}

\begin{figure}[hptb]
\includegraphics[keepaspectratio,angle = -90, width = \columnwidth]{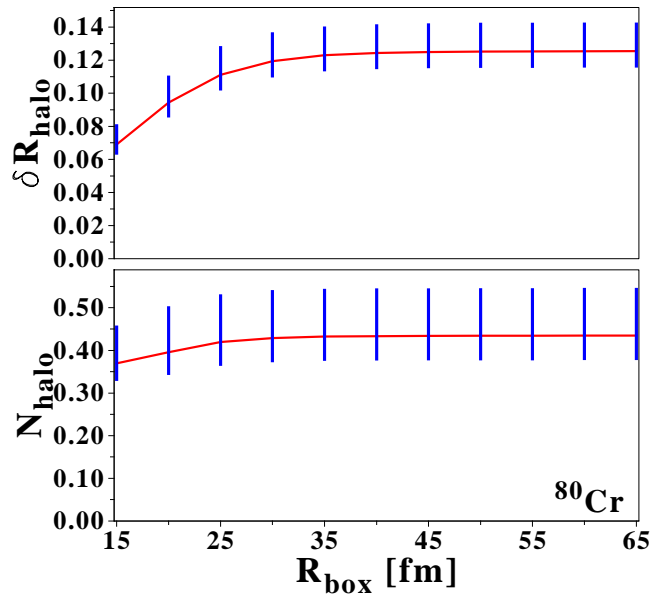}
\caption{(Color Online) Evolution of $N_{\mathrm{halo}}$ and $\delta R_{\mathrm{halo}}$ as a function of the box radius for \mbox{$^{80}$Cr}. }
\label{fig:Cvg_rbox}
\end{figure}

The evolution of the halo factors, which are the most critical observables related to halo properties are represented in \figus\ref{fig:Cvg_rbox} for \mbox{$^{80}$Cr} as a function of $R_{\mathrm{box}}$, for angular-momentum truncations
\mbox{$j_{\mathrm{cut}}^n={65}/{2}$} and \mbox{$j_{\mathrm{cut}}^p={61}/{2}$}, and an energy cutoff \mbox{$E_{\mathrm{max}}=200$}~MeV. The pairing
strength is not refitted for each bin, since the overall effect is found to be negligible.
For small box sizes, $N_{\mathrm{halo}}$ and $\delta R_{\mathrm{halo}}$ are not fully converged and increase with $R_{\mathrm{box}}$, $\delta R_{\mathrm{halo}}$ being the most sensitive quantity. A box radius \mbox{$R_{\mathrm{box}}=40$}~fm, for
which convergence is achieved for all observables of interest, is used in the following.

\subsubsection{Angular-momentum truncation} \label{sec:cvg_jcut}

The choice of $j_{\mathrm{cut}}^n$ and $j_{\mathrm{cut}}^p$ is critical because the partial-wave truncation impacts the way the continuum is
represented in the calculations. Indeed, \mbox{high-$\ell$} orbitals can contribute to nuclear properties such as the pairing gaps. To quantify such an effect, we calculate, in the canonical basis, the
probability distribution of particle $P^{q}_{v^2}(j)$ and pair $P^{q}_{uv}(j)$ occupations as a function of the
single-particle angular-momentum $j$

\begin{eqnarray}
P^{q}_{v^2}(j)&=&\frac{(2j+1)\sum_{n\ell|\ell=j\pm1/2}{v_{nj\ell}^{q}}^2}{\sum_{n'j'\ell'} (2j'+1)\,{v_{n'j'\ell'}^{q}}^2}\,,\\
&&\nonumber\\
P^{q}_{uv}(j)&=&\frac{(2j+1)\sum_{n\ell|\ell=j\pm1/2}u_{nj\ell}^{q}\,v_{nj\ell}^{q}}{\sum_{n'j'\ell'} (2j'+1)\,
u_{n'j'\ell'}^{q}\,v_{n'j'\ell'}^{q}}\,.
\end{eqnarray}

\begin{figure}[hptb]
\includegraphics[keepaspectratio,angle = -90, width = \columnwidth]{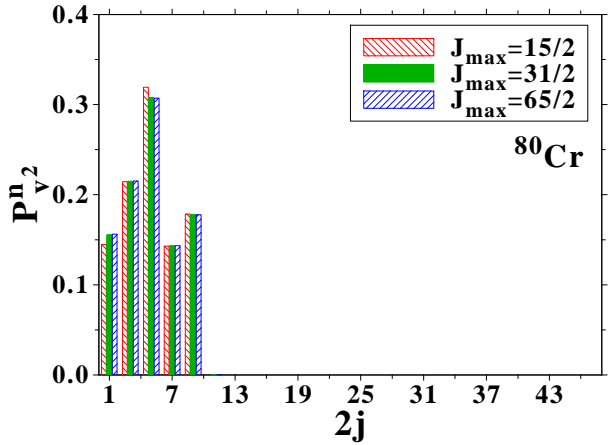}
\caption{(Color Online) Distribution of neutron canonical states occupations $v^{n\,2}$ as fractions of the total strength for
\mbox{$^{80}$Cr} computed with \{SLy4+REN-M\} functionals and different angular momentum truncations
\mbox{$j^n_{\mathrm{cut}}=j^p_{\mathrm{cut}}=J_{\mathrm{max}}$}.\label{fig:pv2}}
\end{figure}

\begin{figure}[hptb]
\includegraphics[keepaspectratio,angle = -90, width = \columnwidth]{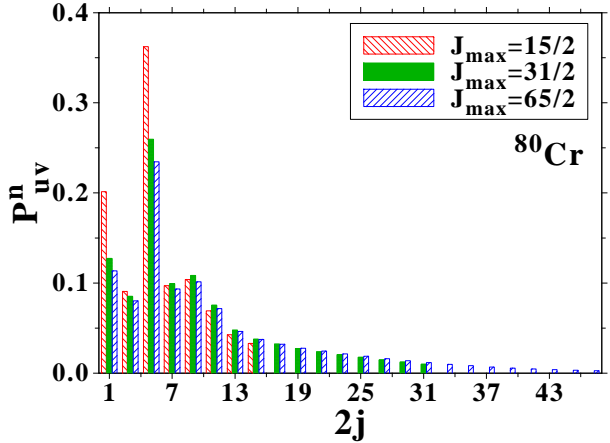}
\caption{(Color Online) Same as \figu\ref{fig:pv2} for the distribution of \mbox{$u^n\,v^n$}.\label{fig:puv}}
\end{figure}

The neutron distribution $P^{n}_{v^2}(j)$ is shown in \figu\ref{fig:pv2} for the halo nucleus \mbox{$^{80}$Cr} and
different angular momentum truncations $j^n_{\mathrm{cut}}=j^p_{\mathrm{cut}}=J_{\mathrm{max}}$. It is found that the $v^{n\,2}$ strength
is mostly distributed over states with \mbox{$j\le9/2$}. As a result, such an occupation distribution is
converged, at least to first approximation, for $j^n_{\mathrm{cut}}=j^p_{\mathrm{cut}}=J_{\mathrm{max}}=15/2$. The corresponding
$P^{n}_{uv}(j)$ distribution (\figu\ref{fig:puv}) extends much further towards high $j$ values. This could be
expected as ${v^q}^2$ is maximum for deeply bound canonical states whereas \mbox{$u^q\,v^q$} is peaked around the Fermi
level and decay slower as the canonical energy increases above $\lambda^{q}$. Correspondingly, the local pair
density $\trho^{q}(r)$ extends further out in space than the normal density $\rho^{q}(r)$~\cite{doba84a}.
As one goes to drip-line (halo) nuclei in particular, $\trho^{q}(r)$ extends very far out and requires many
partial waves to be well represented. This was already highlighted in \refer\cite{matsuo05}. It is clear from
\figu\ref{fig:puv} that a minimal cutoff of $J_{\mathrm{max}}=31/2$ is needed to achieve a reliable enough
description of the pair distribution. One sees in particular that for $J_{\mathrm{max}}=15/2$ (i) the missing $u^n\,v^n$
strength at high $j$ is wrongly redistributed over $j=1/2$ and $j=5/2$, making those states more paired and thus
more localized, whereas (ii) some of the ${v^n}^{2}$ strength of the $j=1/2$ states is transferred to $j=5/2$ states.
Considering that $3s_{1/2}$ and $2d_{5/2}$ states are precisely those building up the halo in $^{80}$Cr, the
latter two effects associated with using a two small angular momentum cutoff inhibits artificially the formation
of the halo. The results are qualitatively identical for the regularization scheme: in both cases, rather high values
of $J_{\mathrm{max}}$ are needed to properly describe the continuum.


\begin{figure}[hptb]
\includegraphics[keepaspectratio,angle = -90, width = \columnwidth]{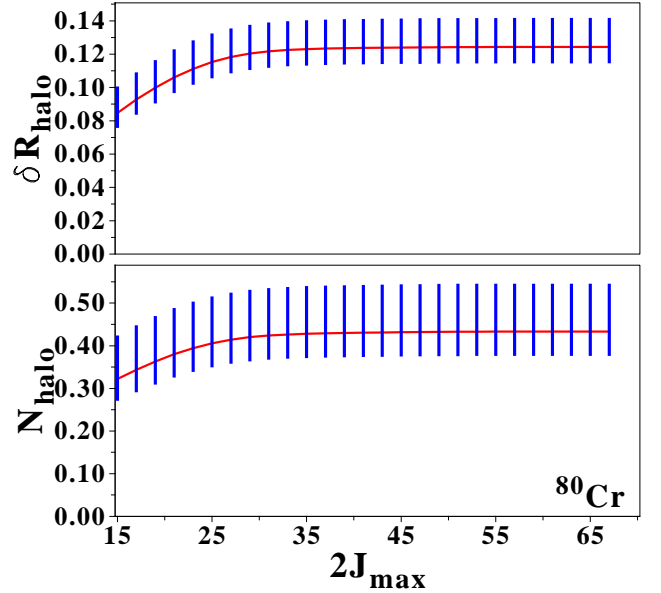}
\caption{(Color Online) Evolution of $N_{\mathrm{halo}}$ and $\delta R_{\mathrm{halo}}$ in \mbox{$^{80}$Cr} and \{SLy4+REN-M\} functionals for
different angular momentum truncations \mbox{$j^n_{\mathrm{cut}}=j^p_{\mathrm{cut}}=J_{\mathrm{max}}$}.\label{fig:halojcut}}
\end{figure}

Indeed, this is what is observed in \figu\ref{fig:halojcut} where $N_{\mathrm{halo}}$ and $\delta R_{\mathrm{halo}}$ are given for
\mbox{$^{80}$Cr} as a function of $J_{\mathrm{max}}$. Both quantities reach a plateau around
\mbox{$J_{\mathrm{max}}=31/2$}.

The previous analysis shows that HFB calculations with too small values of $j_{\mathrm{cut}}^n$ and $j_{\mathrm{cut}}^p$ cannot be trusted
at the limits of stability if one is interested in detailed information
 about potential halos. Of course, considering the ultimate experimental accuracy achievable for matter r.m.s., one
 should not be too extreme as far as the required convergence is concerned. From a theoretical perspective, and
 considering the theoretical error bars on the determination of $N_{\mathrm{halo}}$ and $\delta R_{\mathrm{halo}}$,
 it is necessary to include partial waves up to $j^n_{\mathrm{cut}}=j^p_{\mathrm{cut}}=J_{\mathrm{max}}=31/2$.

\subsubsection{Energy cutoff} \label{sec:cvg_emax}

The value of the energy cutoff $E_{\mathrm{max}}$ in the quasiparticle continuum is an important parameter of the calculation.

\begin{figure}[hptb]
\includegraphics[keepaspectratio,angle = -90, width = \columnwidth]{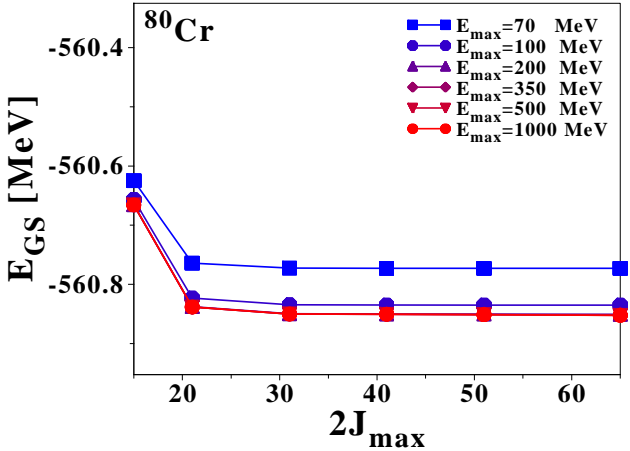}
\caption{(Color Online) Binding energy as a function of the angular momentum cutoff $J_{\mathrm{max}}$ for fixed values of $E_{\mathrm{max}}$ as
obtained from \{SLy4+REG-M\} functional for \mbox{$^{80}$Cr}.\label{fig:Cvg_emax3}}
\end{figure}

\begin{figure}[hptb]
\includegraphics[keepaspectratio,angle = -90, width = \columnwidth]{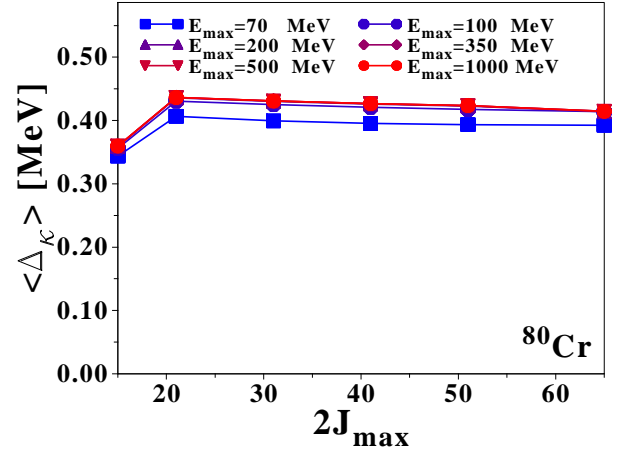}
\caption{(Color Online) Same as in \figu\ref{fig:Cvg_emax3} for the neutron spectral gap.\label{fig:Cvg_emax4}}
\end{figure}

For regularized pairing functionals, the values of $J_{\mathrm{max}}$ and
$E_{\mathrm{max}}$ must be taken large enough that, including a renormalization of the
coupling strength through the re-fitting of data, the observables of interest do not depend on their precise values. It was found that smaller basis truncations could be used for the REG case than for the REN case,
as convergence is reached faster, as exemplified in \figus\ref{fig:Cvg_emax3} and~\ref{fig:Cvg_emax4}.
For the REG case, convergence
for the ground state energy as well as for the neutron pairing gap is almost achieved for $J_{\mathrm{max}}=21/2$ and
$E_{\mathrm{max}}=200~$MeV.

\begin{figure}[hptb]
\includegraphics[keepaspectratio,angle = -90, width = \columnwidth]{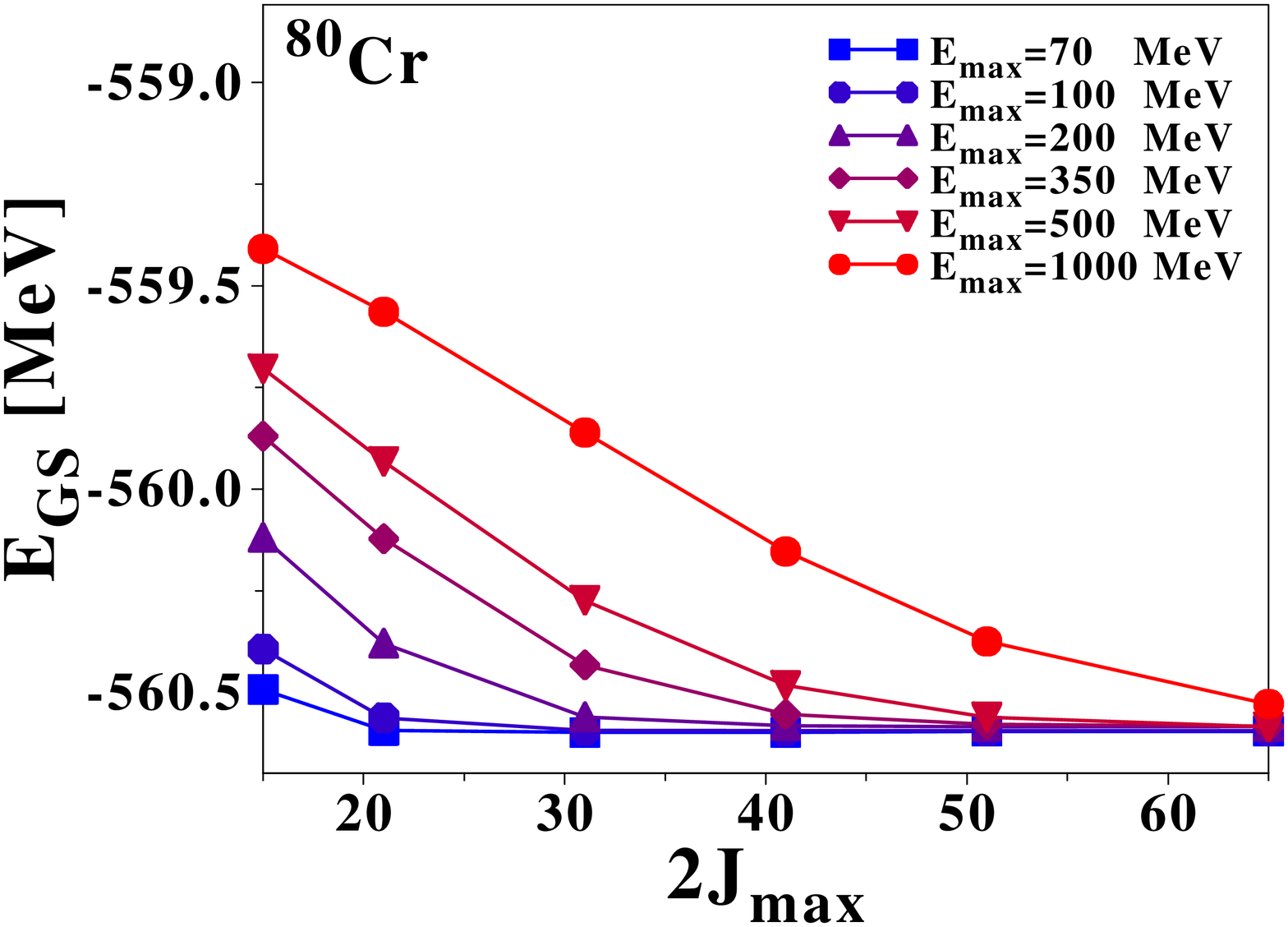}
\caption{(Color Online) Same as in \figu\ref{fig:Cvg_emax3} for the \{SLy4+REN-M\} functional.\label{fig:Cvg_emax1}}
\end{figure}

\begin{figure}[hptb]
\includegraphics[keepaspectratio,angle = -90, width = \columnwidth]{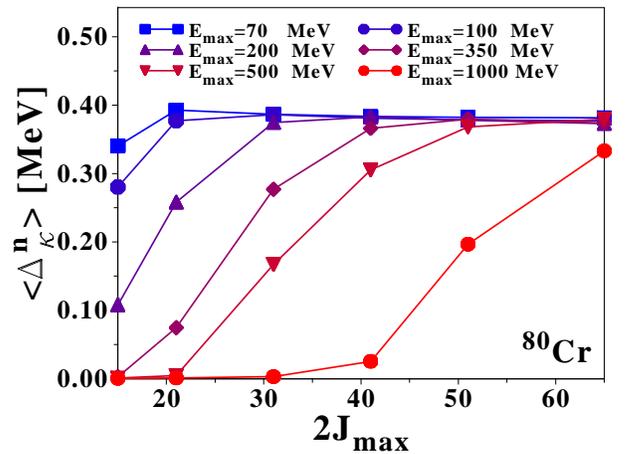}
\caption{(Color Online) Same as in \figu\ref{fig:Cvg_emax4} for the \{SLy4+REN-M\} functional.\label{fig:Cvg_emax2}}
\end{figure}

For renormalized functionals, the situation is more subtle. First, $E_{\mathrm{max}}$ must be large enough for the
result to be independent of its value~\cite{bulgac01}. However, it must be remembered that the field theory
renormalization scheme subtracts a diverging part on the basis that all partial waves below a certain energy cutoff
have been included. Thus, for a given (high enough) $E_{\mathrm{max}}$, the angular momentum truncation must be large enough
to prevent the counter term from removing contributions of states that were not considered in the
first place. This is illustrated in \figus\ref{fig:Cvg_emax1} and~\ref{fig:Cvg_emax2} which display the binding
energy and neutron spectral gap of \mbox{$^{80}$Cr} as a function of the angular momentum cutoff
$J_{\mathrm{max}}$, for fixed values of $E_{\mathrm{max}}$. Note that all $E_{\mathrm{max}}$ values considered are large enough to obtain
converged observables. However, one sees that increasing the energy cutoff necessitates a larger number of
partial waves to reach the converged values for both the energy and the gap. Consequently, it can be counter-productive
to use a safe but unnecessarily large energy cut $E_{\mathrm{max}}$ as it results in the necessity to also
increase $J_{\mathrm{max}}$. On the other hand, the proper description of certain physical phenomena such as halos
intrinsically requires a large number of partial waves as discussed in the previous section. In such a case, one
first fixes $J_{\mathrm{max}}$ and makes sure to use a coherent energy cutoff. In the present work, we use
\mbox{$E_{\mathrm{max}}=200$}~MeV and \mbox{$J_{\mathrm{max}}={65}/{2}$}. This corresponds to an conservative
choice as \mbox{$E_{\mathrm{max}}=70$}~MeV and \mbox{$J_{\mathrm{max}}={41}/{2}$} would be sufficient.

\section{Impact of pairing correlations}
\label{sec:pairing}

\subsection{Pairing anti-halo effect}
\label{sec:antihalo}

In the presence of pairing correlations, the asymptotic of the one-body neutron density takes a different form
from the one it has in the EDF treatment based on an auxiliary Slater determinant~\cite{rotival07a}. Indeed, the
decay constant reads as \mbox{$\kappa^{q}_0=\sqrt{-2m\epsilon^{q}_0/{\hbar^2}}$}, with
\mbox{$|\epsilon^{q}_{0}|=E^{q}_0-\lambda^{q}$}, \mbox{$E^{q}_0={\rm min}_\nu[E^{q}_\nu]$} being the lowest
quasiparticle excitation energy. Considering a canonical state $\phi^q_{0}$ lying at the Fermi level near the
drip-line (\mbox{$e^{q}_0\approx\lambda^{q}\approx 0$}), one finds that \mbox{$E^{q}_{0}\approx\Delta^{q}_0>0$}.
Therefore, in first approximation, paired densities decrease faster than unpaired ones and pairing correlations
induce an \textit{anti-halo effect} by localizing the one-body density~\cite{bennaceur99,bennaceur00,yamagami05}.

\begin{figure}[hptb]
\includegraphics[keepaspectratio, angle = -90, width = \columnwidth]{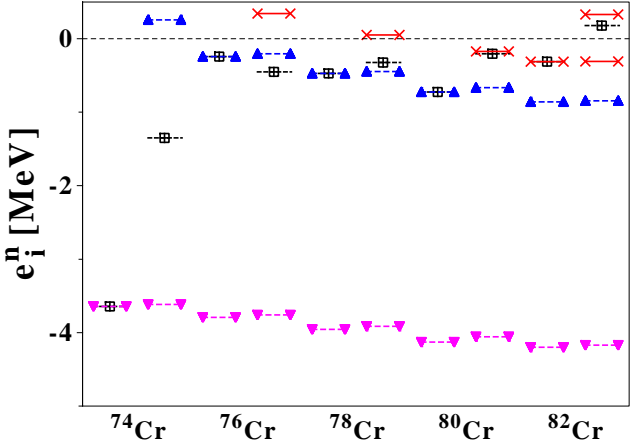}
\caption{ \label{fig:Cr_HF} (Color Online) Neutron single-particle energies $e^n_i$ along the Cr isotopic chain. Left: eigenvalues of $h^{n}$ obtained when omitting pairing correlations in the calculation. The non-resonant continuum spectrum has been omitted while resonant states with positive $e^n_i$ are drawn as discrete states for convenience. Right: canonical energies obtained including pairing correlations through a
mixed-type REG-M functional. The Skyrme SLy4 functional is used in both cases. The conventions used for single-particle states labeling are identical to
\refer\cite{rotival07a} and are recalled in \figu\ref{fig:ref}. The Fermi level \mbox{($\mathbf{\boxplus}$)} is set
to the last filled orbital for the EDF treatment based on an auxiliary Slater determinant.}
\end{figure}

\begin{figure}[hptb]
\includegraphics[keepaspectratio, angle = -90, width = \columnwidth]{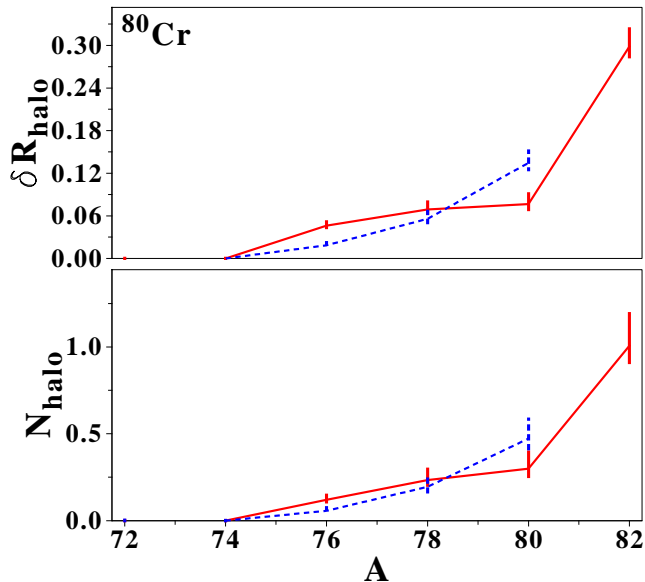}
\caption{ \label{fig:Cr_HF2} (Color Online) Halo criteria $N_{\mathrm{halo}}$ and $\delta R_{\mathrm{halo}}$ for Cr isotopes, obtained through spherical HF (solid lines) and HFB (dashed lines)
calculations with the Skyrme SLy4 functional. For HFB calculations, mixed-type REG-M pairing is used. }
\end{figure}

To evaluate the quantitative impact of this effect, drip-line Cr isotopes have been calculated with and without
explicit treatment of pairing correlations. In the latter case, a {\it filling approximation}~\cite{PerezMartin:2008yv} is used for
incomplete spherical shells. In both cases, the SLy4 Skyrme functional is used. When including pairing, a
mixed-type pairing is added. A comparison between neutrons single-particle levels is represented in
\figu\ref{fig:Cr_HF} for the last bound nuclei, \mbox{$^{82}$Cr} being predicted to be bound when pairing
correlations are excluded from the treatment. This is interesting in itself as it shows that pairing correlations
can change the position of the drip-line and modify in this way the number of halo candidates over the nuclear
chart. There is only little difference between the canonical energies $e^n_i$ in the two cases\footnoteb{The canonical basis is identical to the eigenbasis of $h$ is the zero-pairing limit.}. However, the values
of the halo criteria $N_{\mathrm{halo}}$ and $\delta R_{\mathrm{halo}}$ are significantly different, i.e. the neutron halo is
significantly quenched in \mbox{$^{80}$Cr} when pairing is omitted whereas the situation is reversed in the
lighter isotopes, as seen in \figu\ref{fig:Cr_HF2}.
\par

\begin{figure}[hptb]
\includegraphics[keepaspectratio, angle = -90, width = 0.8\columnwidth]{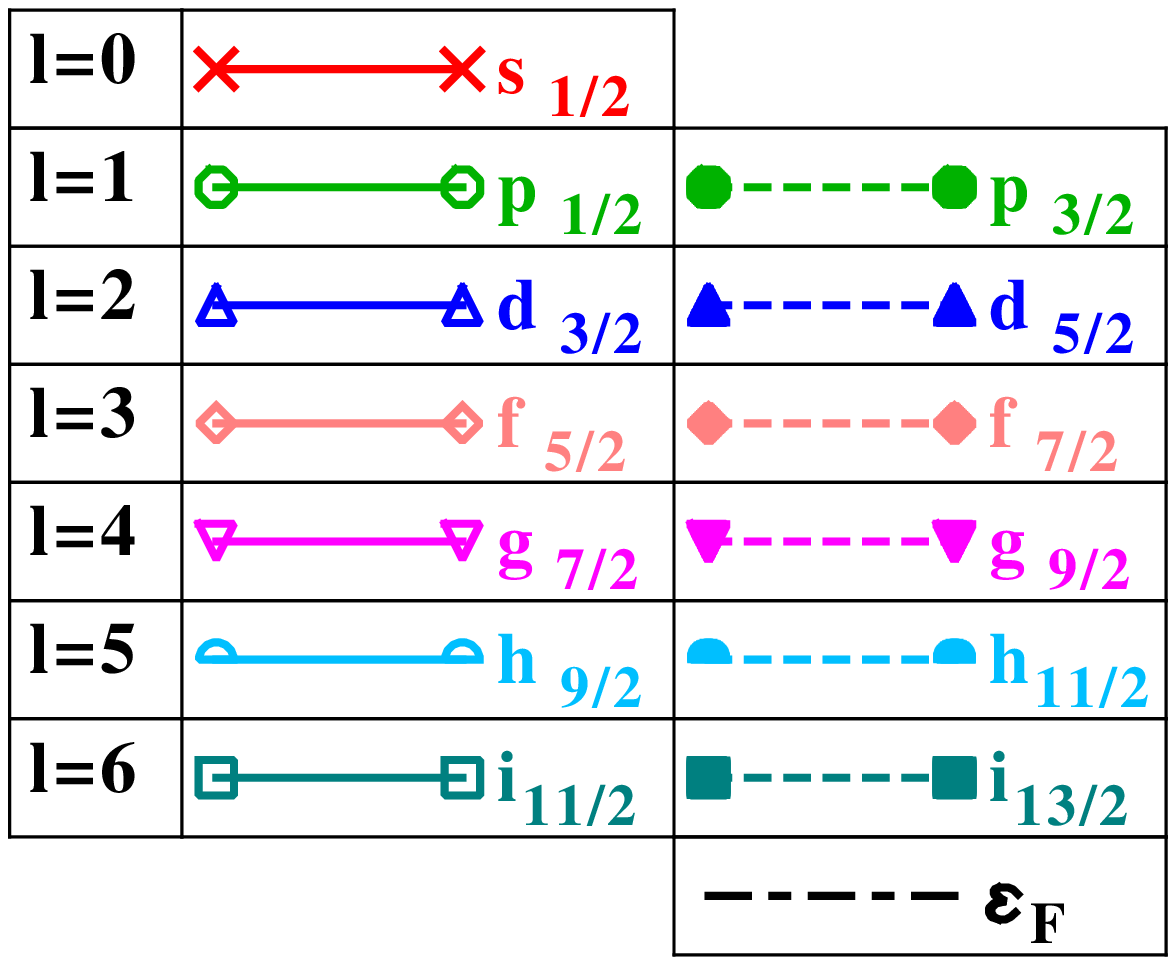}
\caption{\label{fig:ref} (Color Online) Conventions used in all the figures for the labeling of individual states and of the
chemical potential.}
\end{figure}

Such results underline that pairing correlations affect halos in two opposite ways. Pairing (i) inhibits the
formation of halos through the anti-halo effect (ii) enhance the formation of halos by scattering nucleons to less
bound states with smaller decay constants. For example, the anti-halo effect dominates in $^{76}$Cr and $^{78}$Cr
whereas the promotion of neutrons into the weakly bound $3s_{1/2}$ makes the halo to be more pronounced in
\mbox{$^{80}$Cr} when pairing is included.

\begin{figure}[hptb]
\includegraphics[keepaspectratio, angle = -90, width = \columnwidth]{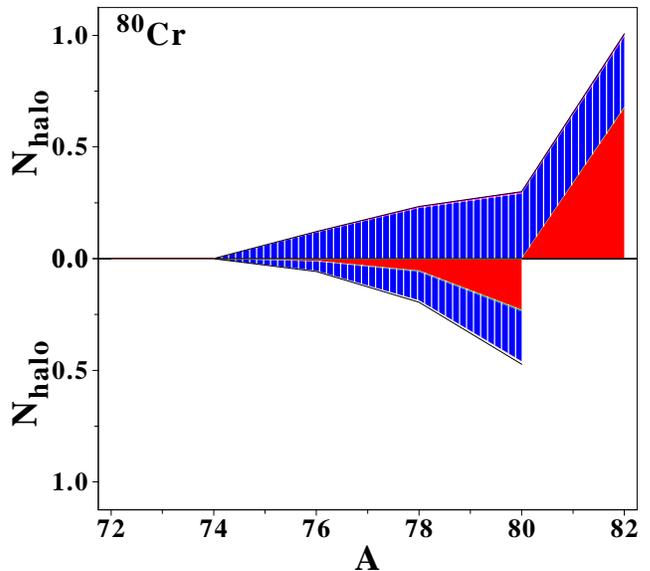}
\caption{ \label{fig:Cr_HF3} (Color Online) Contributions of individual orbital to the halo: EDF calculation of the least bound
Cr isotopes with (bottom) and without (top) pairing. The $2d_{5/2}$ orbital corresponds to vertically-dashed
areas, and $3s_{1/2}$ orbital to filled ones. The other contributions are found to be negligible.}
\end{figure}

When pairing is omitted, the number of nucleons in the $2d_{5/2}$ orbital increases linearly as one goes from
$^{76}$Cr to $^{80}$Cr. At the same time, the $2d_{5/2}$ shell becomes more bound as a result of self-consistency
effects. This explains the negative curvature of $\delta R_{\mathrm{halo}}$ as one goes from $^{76}$Cr to $^{80}$Cr, while
$N_{\mathrm{halo}}$ is indeed almost linear. In \mbox{$^{82}$Cr}, two effects contribute to the very pronounced halo that
is predicted in the absence of pairing correlations (i) the $3s_{1/2}$ gets fully occupied whereas (ii) the
pairing anti-halo effect is inoperative. The results discussed above are further illustrated in
\figu\ref{fig:Cr_HF3} where the contributions of different single-particle states to the halo are shown.

\begin{table}[htbp]
\setlength{\extrarowheight}{2pt}
\begin{tabular}{|l||c|c|c|}
\hline & \multicolumn{3}{c|}{$^{80}$Cr (pairing)}\\
\hline $N_{\mathrm{halo}}$ & \multicolumn{3}{c|}{$0.473$}\\
\hline & $e^n_i$~[MeV] & ${v_{i\,2}^n}$ & $N_{\mathrm{halo,i}}$ \\
\hline
 & $>0.0$    & ---             & $\sim 1.3.10^{-2}$\\
${3s_{1/2}}$ & ${-0.173}$  & ${0.451}$   & ${0.225}$  \\
${2d_{5/2}}$ & ${-0.665}$  & ${0.828}$   & ${0.222}$ \\
$1g_{9/2}$ & $-4.056$ & $0.993$          & $0.008$          \\
$1f_{5/2}$ & $-8.673$ & $0.999$          & $0.001$          \\
$2p_{1/2}$ & $-8.920$ & $0.999$          & $0.002$          \\
$Other$ & $<-10.0$    & ---             & $\sim 2.8\times10^{-4}$\\
\hline \hline
& \multicolumn{3}{c|}{$^{82}$Cr (no pairing)}\\
\hline $N_{\mathrm{halo}}$ & \multicolumn{3}{c|}{$1.007$}\\
\hline & $e^n_i$~[MeV]  & ${v_{i\,2}^n}$ & $N_{\mathrm{halo,i}}$ \\
\hline
 & $>0.0$    & ---             & $0.000$\\
$3s_{1/2}$ & $-0.312$ & $1.000$ & $0.675$     \\
$2d_{5/2}$ & $-0.858$ & $1.000$ & $0.314$      \\
$1g_{9/2}$ & $-4.200$ & $1.000$ & $0.010$      \\
$1f_{5/2}$ & $-8.783$ & $1.000$ & $0.001$       \\
$2p_{1/2}$ & $-9.078$ & $1.000$ & $0.002$       \\
$Other$ & $<-10.0$    & ---    & $\sim 2.7\times10^{-3}$ \\
\hline \hline
& \multicolumn{3}{c|}{$^{80}$Cr$^\ast$ (pairing)}\\
\hline $N_{\mathrm{halo}}$ & \multicolumn{3}{c|}{$0.582$}\\
\hline & $e^n_i$~[MeV]  & ${v_{i\,2}^n}$ & $N_{\mathrm{halo,i}}$ \\
\hline
 & $>0.0$    & ---             & ??????\\
$3s_{1/2}$ & $-0.312$ & $0.451$          & $0.305$           \\
${2d_{5/2}}$   & ${-0.858}$  & ${0.828}$ &  ${0.260}$ \\
$1g_{9/2}$ & $-4.200$ & $0.993$          & $0.010$            \\
$1f_{5/2}$ & $-8.784$ & $0.999$          & $0.002$            \\
$2p_{1/2}$ & $-9.078$ & $0.999$          & $0.002$          \\
$Other$ & $<-10.0$    & ---             & $\sim 2.7\times10^{-3}$ \\
\hline
\end{tabular}
\caption{\label{tab:Cr_HF} $N_{\mathrm{halo}}$, individual contributions to the halo $N_{\mathrm{halo,i}}$ (see Paper I for the
definition of $N_{\mathrm{halo,i}}$), single-particle canonical energies $e^n_i$ and occupations ${v^{n\,2}_i}$ for (i)
\mbox{$^{80}$Cr} with pairing (ii) \mbox{$^{82}$Cr} without pairing, and (iii) \mbox{$^{80}$Cr$^\ast$} with
pairing but using the single particle spectrum calculated in the absence of pairing.}
\end{table}

To specifically extract the pairing anti-halo effect, we now perform a toy model calculation of a fictitious
\mbox{$^{80}$Cr$^\ast$} nucleus. Filling the single-particle wave-functions obtained from the calculation of
\mbox{$^{82}$Cr} without pairing correlations with the canonical occupations obtained from the
calculation of \mbox{$^{80}$Cr} with pairing, we extract $N_{\mathrm{halo,i}}$ from each orbital\footnoteb{Single-particle states extracted from the calculation of \mbox{$^{80}$Cr} without pairing cannot be used because the
essential $3s_{1/2}$ orbital belongs to the continuum and has plane wave asymptotic in this case.}. Such a
procedure allows one to isolate, in semi-quantitative manner, the net change of $N_{\mathrm{halo}}$ due to the difference in
the asymptotic of paired orbitals keeping their occupation fixed.
Doing so, it is found that the contribution of the $3s_{1/2}$ state to the halo is smaller in \mbox{$^{80}$Cr}
than in \mbox{$^{80}$Cr$^\ast$} by about $26\,\%$. For the $2d_{5/2}$ orbital the suppression of $N_{\mathrm{halo,i}}$ is
about $15\,\%$ (the value of $r_0$ is also different for \mbox{$^{80}$Cr} and \mbox{$^{80}$Cr$^\ast$}, but
this does not affect the results significantly). As a result, one sees that if the anti-halo effect were
ineffective, the scattering of particles into higher-lying orbitals would bring $N_{\mathrm{halo}}$ from $0.300$ to
$0.582$. Instead, $N_{\mathrm{halo}}$ is only increased to $0.473$ in the full fledged calculation of $^{80}$Cr with
pairing correlations, i.e. the anti-halo effect reduces the potential increase by $40\,\%$.

To conclude, there is no simple answer as to whether pairing correlations enhance or hinder the formation of
halos. The net result depends on structure details of the particular nucleus of interest~\cite{grasso01a,grasso06}.

\subsection{Decorrelation from the pairing field}
\label{sec:hamamoto}

An additional effect might come into play as far as the role of pairing in the formation of halos is concerned.
Very weakly bound \mbox{$\ell=0$} orbitals are expected to decouple from the pairing field as the neutron
separation energy goes to zero~\cite{hamamoto03,hamamoto04,hamamoto05}. As a result, such an orbital would not be
subject to the anti-halo effect and may develop a very long tail.

\begin{figure}[hptb]
\includegraphics[keepaspectratio, angle = -90, width = \columnwidth]%
{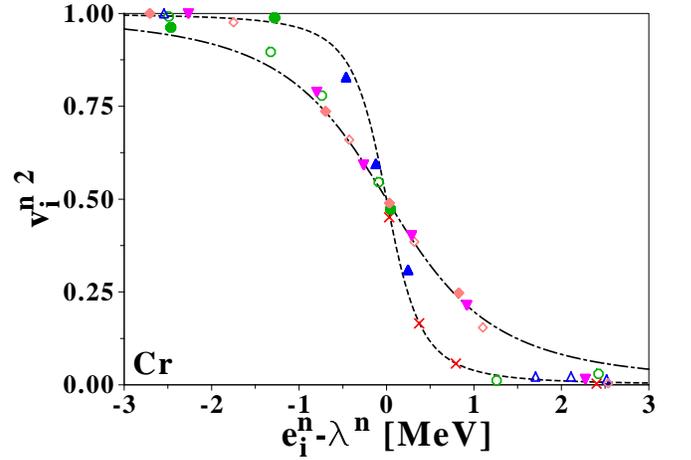} \caption{ \label{fig:Cr_occ} (Color Online) Neutron canonical occupations $v^{n\,2}_i$ as a function of
\mbox{$(e^{n}_i-\lambda^{n})$} for all drip-line Cr isotopes. The calculations are done with the \{SLY4+REG-M\}
functional. The profiles computed from the BCS-type formul\ae~ using the minimum (dashed line) and maximum
(dashed-dotted line) spectral neutron gaps \mbox{$\langle\Delta^n_\kappa\rangle$} among all those isotopes are shown
for reference. Conventions for labeling individual states are given in \figu\ref{fig:ref}.}
\end{figure}

The signature of this effect can be seen in the single-particle occupation profile. The canonical occupations of
all neutron single-particle states in all drip-line Cr isotopes are gathered in \figu\ref{fig:Cr_occ} and plotted
as a function of \mbox{$(e^{n}_i-\lambda^{n})$}, where $e^{n}_i$ is the neutron canonical energy and $\lambda^{n}$
the neutron Fermi level. Those occupations are compared to the BCS formula
\begin{equation}
v^{n\,2}_i=\frac{1}{2}\left(1-\frac{e^{n}_i-\lambda^{n}}{\sqrt{(e^{n}_i-\lambda^{n})^2 + \langle \Delta^{n}_{\kappa}
\rangle^2}}\right)\,, \label{eq:BCS}
\end{equation}
calculated using the maximum/minimum neutron spectral gap $\langle \Delta^{n}_{\kappa} \rangle$ found among all
drip-line Cr isotopes. The calculations have been performed using the \{SLY4+REG-M\} functional.

\begin{figure}[hptb]
\includegraphics[keepaspectratio, angle = -90, width = \columnwidth]{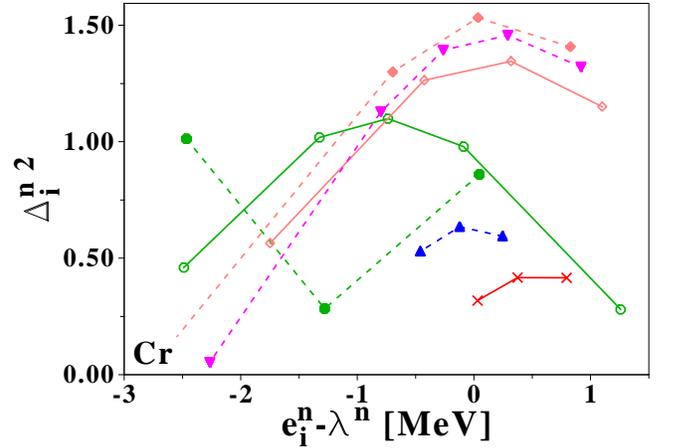}
\caption{ \label{fig:Cr_occ2} (Color Online) Individual neutron canonical gaps $\Delta^{n}_i$ computed for drip-line Cr isotopes with the
\{SLY4+REG-M\} functional and plotted as a function of $(e^{n}_i-\lambda^{n})$. Conventions for individual states
are given in \figu\ref{fig:ref}.}
\end{figure}

The \mbox{$s$-wave} occupation probability follows closely the BCS-type profile calculated using the minimal
spectral gap. The high-$\ell$ orbitals follow well the BCS-type profile computed with the large spectral gap. This
corroborates the trend discussed in \refers\cite{hamamoto03,hamamoto04,hamamoto05} and underlines that $\ell=0$
orbitals with $e^{n}_i\approx\lambda^{n}\approx0$ are less paired than high-$\ell$ ones. This is also confirmed by
looking at individual gaps given in \figu\ref{fig:Cr_occ2}. The $3s_{1/2}$ state displays a smaller gap than other
orbitals as it approaches the Fermi level (from above), the latter reaching the continuum threshold.

On the other hand, the canonical gap of the $3s_{1/2}$ state remains significant as it crosses the Fermi level and
the anti-halo effect is still in effect for that orbital. In fact, the critical decoupling of $s$ orbitals from
the pairing field, discussed in \refers\cite{hamamoto03,hamamoto04,hamamoto05} through schematic HFB
calculations, becomes fully effective only when the Fermi level and the $s$ state energy are both of the order of
a few keVs. Such extreme situations (i) are not reached in realistic systems~\cite{hamamotopriv} (ii) would
require an accuracy on the predicted value of the separation energy which is far beyond present capacities of
EDF calculations.

\subsection{Importance of low densities}
\label{sec:pair_loc}

The values of $\eta$ and $\alpha$ entering the pairing functional strongly affect the spatial localization of the
pairing field, and thus the gaps of weakly-bound orbitals lying at the
nuclear surface.

In previous studies~\cite{doba01b}, it was found that the size of a neutron halo could change by one order of
magnitude when the pairing functional evolves from a volume to a extreme-surface type. However (i) the evaluation
of the halo size was performed through the Helm model, the limitations of which have been pointed out in Ref.~\cite{rotival07a},
and (ii) the standard regularization scheme was used with extreme-surface pairing functionals which, as discussed
in \sect\ref{sec:pathos_finite}, leads to unsafe predictions of halo properties. The renormalization scheme is used in
the present section as it prevents any un-physical feature from appearing with extreme surface pairing
functionals. The properties of the last bound Cr isotopes are evaluated for different pairing functionals: (i)
\mbox{$\alpha=1$} and \mbox{$\eta\in[0,1]$}, and (ii) \mbox{$\eta=1$} and \mbox{$\alpha\in[1,0.1]$}.

\begin{figure}[hptb]
\includegraphics[keepaspectratio, angle = -90, width = \columnwidth]%
{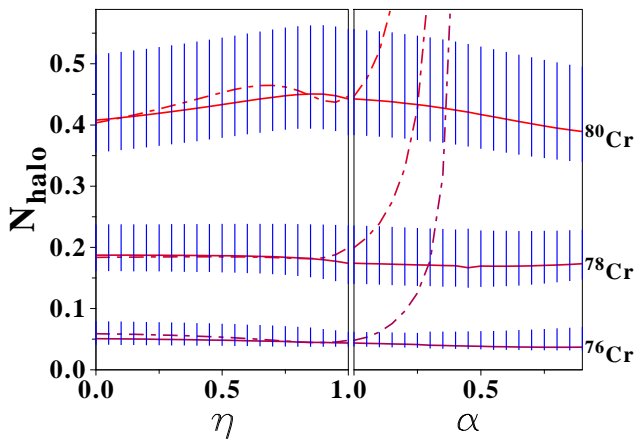} \caption{(Color Online) $N_{\mathrm{halo}}$ parameter for Cr isotopes calculated using
renormalized REN-X pairing functionals with different form factors (see text). Results for regularized REG-X
functionals diverge for very strong surface pairing correlations and are represented in dashed
lines.\label{fig:SV_RDFTX_N}}
\end{figure}
\begin{figure}[hptb]
\includegraphics[keepaspectratio, angle = -90, width = \columnwidth]%
{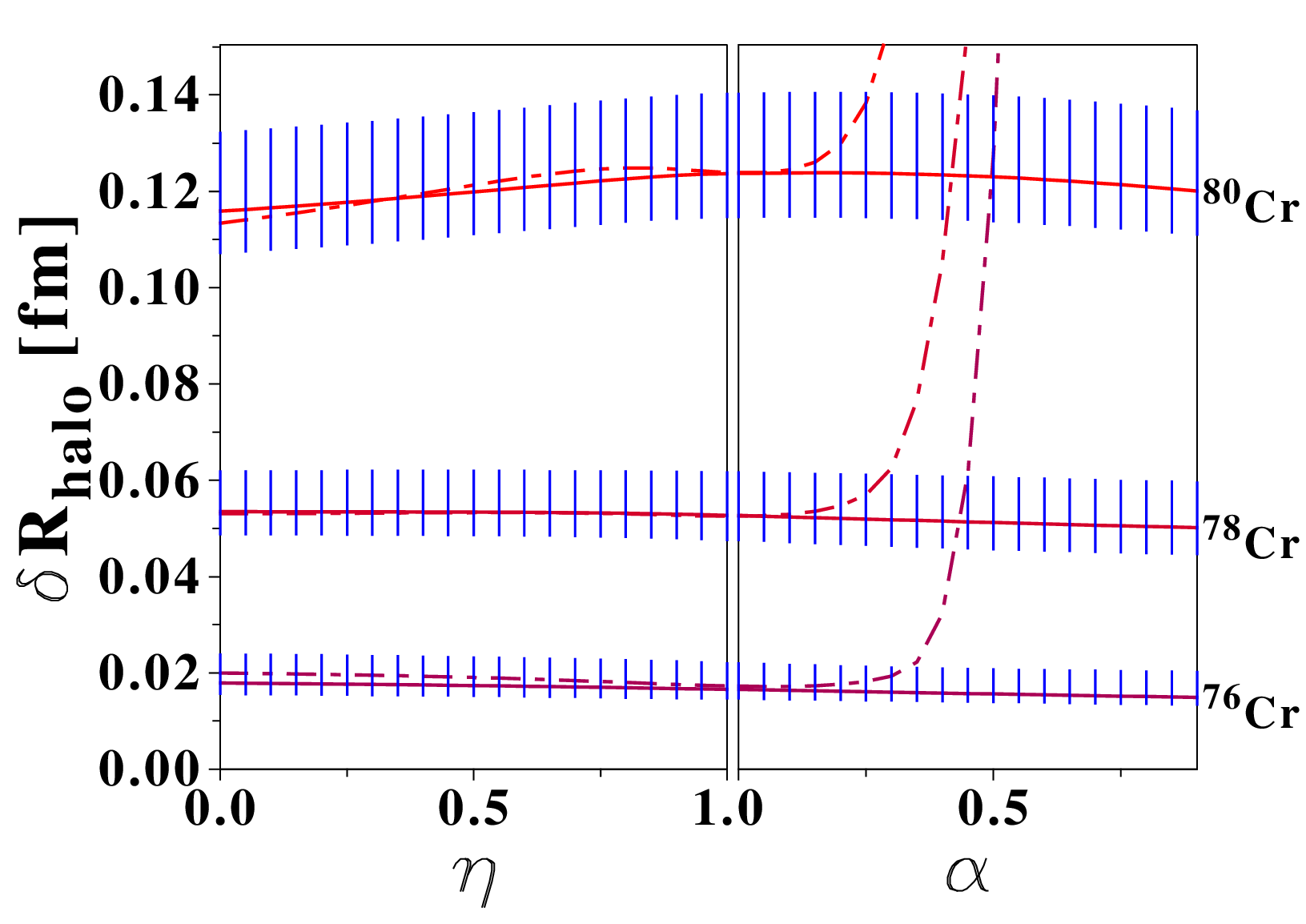} \caption{(Color Online) Same as \figu\ref{fig:SV_RDFTX_N} for $\delta
R_{\mathrm{halo}}$.\label{fig:SV_RDFTX_R}}
\end{figure}

\begin{figure}[hptb]
\includegraphics[keepaspectratio, angle = -90, width = \columnwidth]%
{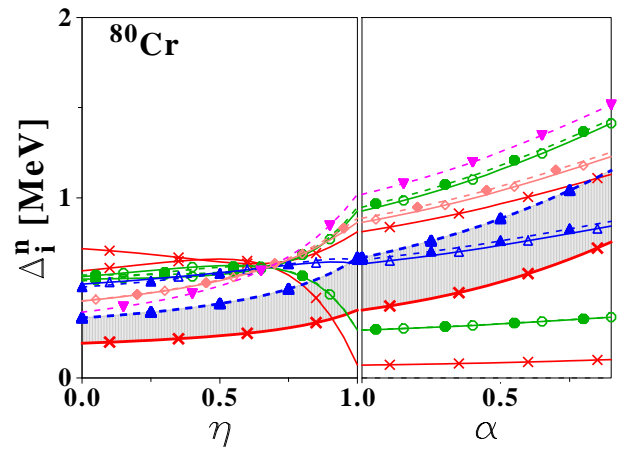}
\caption{(Color Online) Neutron single-particle pairing gaps $\Delta^q_i$ for \mbox{$^{80}$Cr} and different REN-X pairing functionals. Conventions
for labeling individual states are given in \figu\ref{fig:ref}.\label{fig:Cr80_gap}}
\end{figure}

\begin{figure}[hptb]
\includegraphics[keepaspectratio, angle = -90, width = \columnwidth]%
{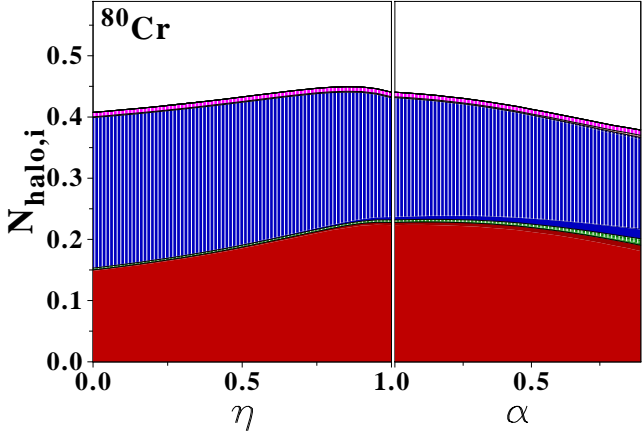}
\caption{(Color Online) Individual neutron contributions $N_{\mathrm{halo,i}}$ for \mbox{$^{80}$Cr} and different renormalized REN-X pairing functionals.
The contribution from the $2d_{5/2}$ orbital corresponds to vertically-dashed areas, and the one from the $3s_{1/2}$ state
to filled ones.\label{fig:Cr80_RDFTX_decomp}}
\end{figure}

Overall, neutron canonical energies evolve very little with $(\eta,\alpha)$. The evolution of canonical
pairing gaps $\Delta^{n}_i$ with $(\eta,\alpha)$ is presented in \figu\ref{fig:Cr80_gap}. For surface-enhanced pairing functionals,
well-bound orbitals residing in the center of the nucleus become less paired. On the other hand, pairing gaps of
states close to the Fermi level increase as the effective pairing strength becomes more important at the nuclear
surface\footnoteb{Keeping the neutron spectral gap unchanged in
\mbox{$^{120}$Sn}, and as some orbitals become less paired, others display larger individual gaps.}. Considering the theoretical error bars, the values of $N_{\mathrm{halo}}$ and $\delta R_{\mathrm{halo}}$
presented in \figus\ref{fig:SV_RDFTX_N} and \ref{fig:SV_RDFTX_R} can be considered to be almost independent of the
density-dependent form factor of the pairing functional, although the anti-halo effect becomes more effective as
$\alpha$ decreases.
 $N_{\mathrm{halo}}$ and $\delta R_{\mathrm{halo}}$ are maximal for the standard surface pairing functional
\mbox{$\eta=\alpha=1$}, for which the occupation of the $3s_{1/2}$ state and its localization due to an increased
coupling to the pairing field presents the most favorable compromise for the halo to develop. This reflects on the
composition of the neutron halo in \mbox{$^{80}$Cr} displayed in \figu\ref{fig:Cr80_RDFTX_decomp}. Finally, results obtained with the regularization scheme are also plotted in \figus\ref{fig:SV_RDFTX_N} and \ref{fig:SV_RDFTX_R} for comparison. As long as \mbox{$\alpha=1$}, $N_{\mathrm{halo}}$ and $\delta R_{\mathrm{halo}}$ are almost identical for the
regularization and renormalization schemes, which proves that results presented in Ref.~\cite{rotival07a} using the regularized
scheme are valid and that both REG-X or REN-X functionals can be used safely with ``standard'' density dependencies.

As a conclusion, the impact of the low density characteristics of the pairing functional on halo properties is
found to be small, as long as the adequate renormalization scheme is used to prevent the formation of the
un-physical gas of bound di-neutrons. Consequently, experimental constraints on pairing
localization and the effective pairing strength based solely on halo properties are unlikely.

\section{Large-scale predictions}
\label{sec:fullscale}

We now turn to predictions of halo nuclei over the nuclear chart. In addition, the
dependence of such predictions on some of the characteristics of the Skyrme functional used in the calculations
are briefly investigated.

\subsection{Results for different particle-hole functionals}
\label{sec:ph}

To study the impact of the particle-hole functional characteristics on the formation of halos, $N_{\mathrm{halo}}$
and $\delta R_{\mathrm{halo}}$ have been computed for Cr isotopes using the set of functionals presented in
\sect\ref{sec:skyrme_ph} combined with a mixed-type REG-M pairing functional. The results obtained with SLy4 and
presented in Ref.~\cite{rotival07a} act as a reference point.

\begin{figure*}
\centering
\includegraphics[keepaspectratio,angle = -90, width = 0.7\textwidth]%
{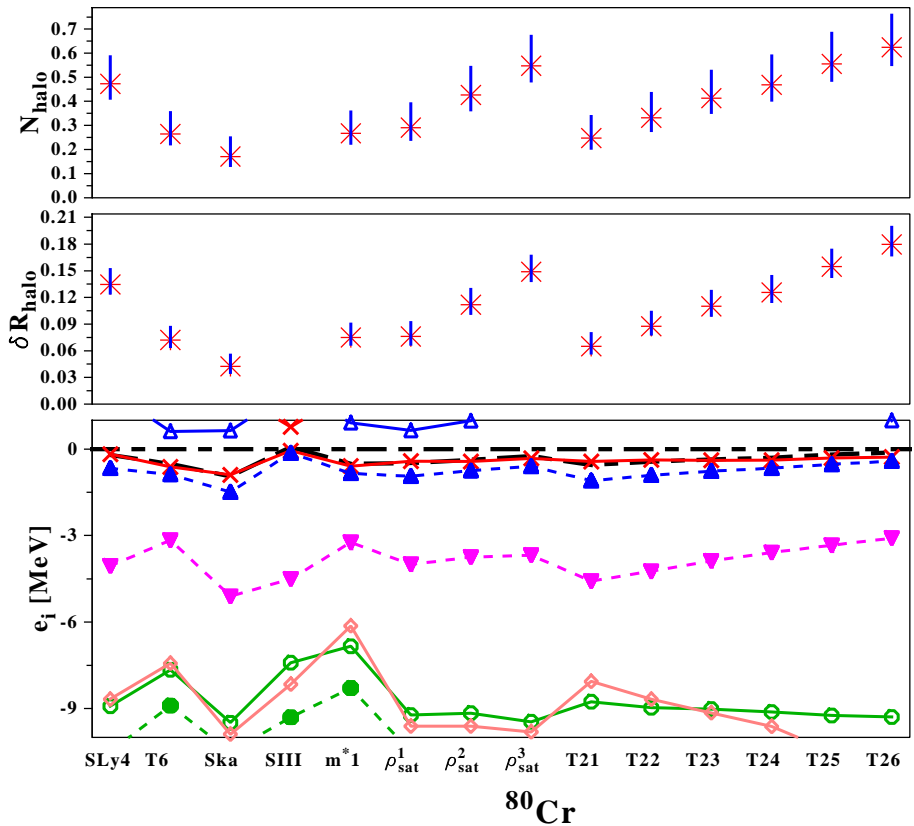} \caption{(Color Online) Single-particle spectrum and halo properties for \mbox{$^{80}$Cr},
computed with different particle-hole functionals (see text) and mixed-type pairing. Conventions
for labeling individual states are given in \figu\ref{fig:ref}.} \label{fig:comp_skyrme80}
\end{figure*}

The halo parameters $N_{\mathrm{halo}}$ and $\delta R_{\mathrm{halo}}$ as well as neutron canonical energies are
displayed in \figu\ref{fig:comp_skyrme80} for \mbox{$^{80}$Cr}. We observe that

\begin{figure}[hptb]
{\includegraphics[keepaspectratio,angle = -90,width=\columnwidth]%
{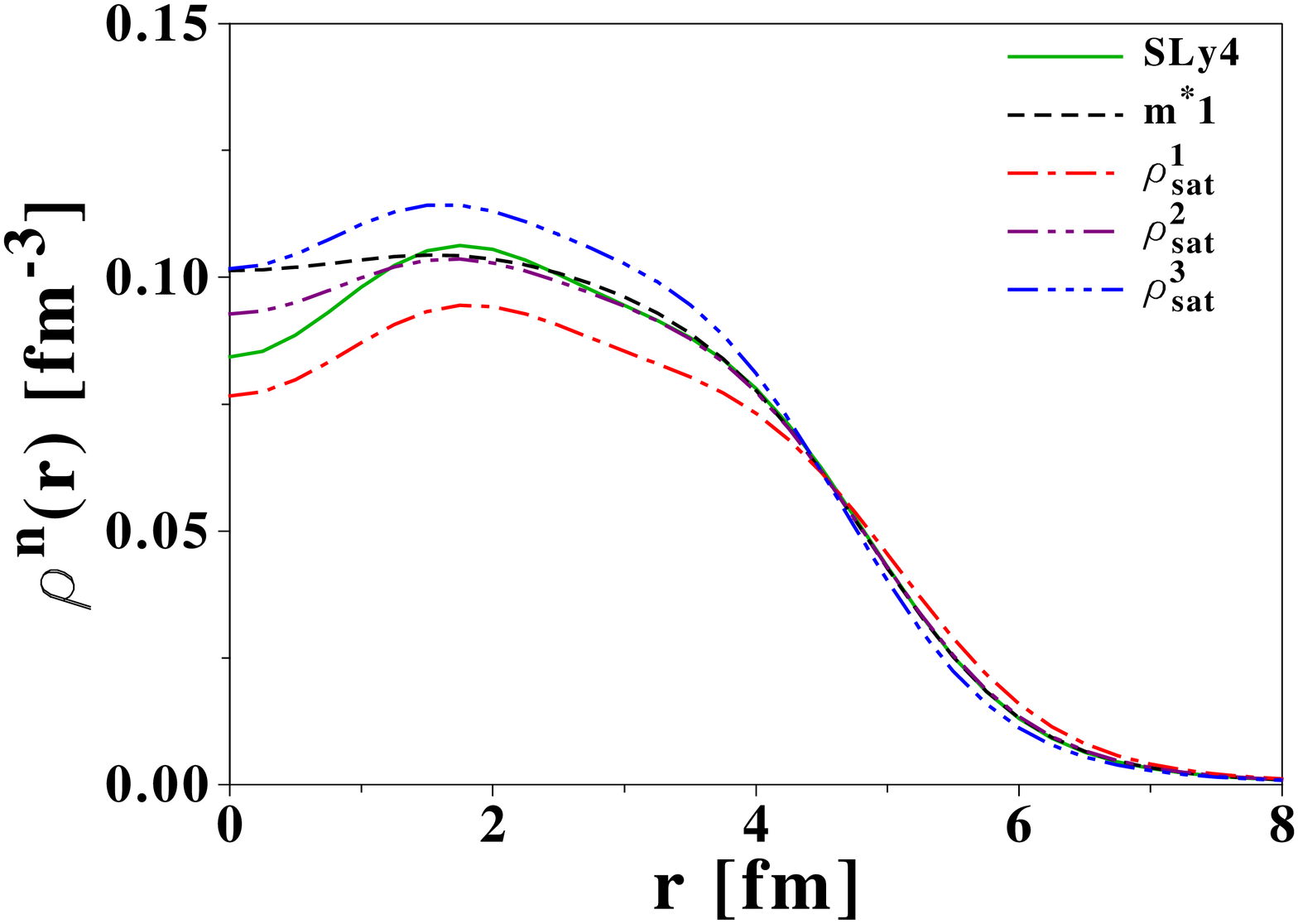}\vspace{10pt} }\\
{\includegraphics[keepaspectratio,angle = -90,width=\columnwidth]%
{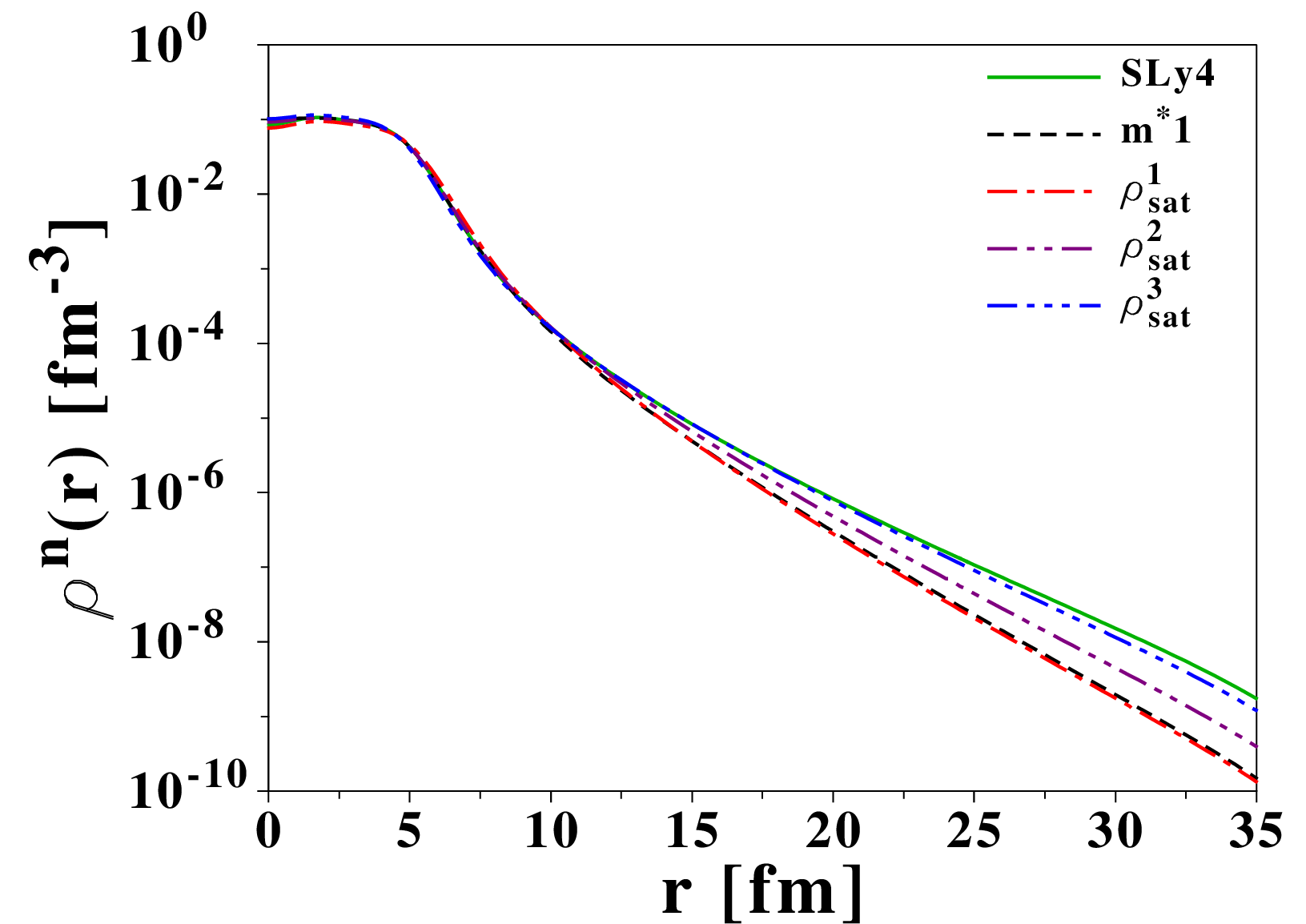} } \caption{ \label{fig:Cr_rhosat} (Color Online) Neutron densities for \mbox{$^{80}$Cr} (top: linear scale,
bottom: logarithmic scale) with the \{SLy4+REG-M\} functional (reference), and refits of the Skyrme parameters
corresponding to (i) a different isoscalar effective mass \mbox{$m^\ast=0.70\rightarrow1.00$}, and (ii) different
nuclear saturation densities: \mbox{$\rho^{1/2/3}_{\mathrm{sat}}=0.145\,/\,0.160\,/\,0.175$~fm$^{-3}$}.}
\end{figure}

\begin{itemize}
\item Skyrme functionals with an isoscalar effective mass equal to one ($m^\ast1$ and T6) predict denser single-particle
spectra around the Fermi level. As a consequence, the \mbox{$2d_{5/2}-1g_{9/2}$} energy gap corresponding to the
core excitation energy scale $E'$ is reduced. Additionally, the Fermi level is shifted down, which enhances the
separation energy $E$. Both effects contribute to hindering the halo formation.
\item The modification of $K_\infty$ affects collective properties, such as the breathing mode energy~\cite{blaizot95}.
It also impacts the canonical spectrum. As a result, \mbox{$^{80}$Cr} is predicted to be unbound with the
\{SIII+REG-M\} parametrization which has a particularly large incompressibility $K_\infty=355$~MeV.
\item As the nuclear matter saturation density increases (from $\rho_{\mathrm{sat}}^1$ to $\rho_{\mathrm{sat}}^3$),
the nuclear interior becomes denser, as shown in \figu\ref{fig:Cr_rhosat}. Through self-consistency, a denser
nuclear interior generates a sharper surface, that ultimately makes weakly bound orbitals to be less coupled to
the nuclear potential and thus even less bound. As a result, the density extends further out asymptotically. While
the \mbox{$2d_{5/2}-1g_{9/2}$} energy gap remains the same, the tail excitation energy $E$ decreases with $\rho_{\mathrm{sat}}$
and the halo factors increase.
\item Tensor couplings may also impact the formation of halos; as it was found
for light systems~\cite{myo07}. For medium-mass nuclei, a series of recent studies have assessed the impact of
tensor couplings on the evolution of spherical single-particle
shells~\cite{otsuka06,doba06,brown06,colo07,lesinski07a}. To study the impact on halos, newly available Skyrme
T21-T26 functionals have been used~\cite{lesinski07a}. This particular series of parameterizations differ through their like-particle tensor coupling and an increasing spin-orbit coupling as one goes from T21 to T26. The like-particle coupling constant
is (i) negative for T21, which corresponds to a repulsion between particles of identical isospin, (ii) zero for
the functional T22, which makes it similar to SLy4, and (iii) increasingly positive from T23 to T26. All TXX
functionals display otherwise the same infinite matter properties, as seen in \tab\ref{tab:skyrme_prop}: the
variation of single-particle properties are solely due to the tensor and spin-orbit interactions. The results are shown in
\figu\ref{fig:comp_skyrme80}. As the like-particle tensor interaction becomes attractive, both \mbox{$3s_{1/2}-2d_{5/2}$}
and \mbox{$2d_{5/2}-1g_{9/2}$} energy gaps involving neutron states of the same parity decrease. The Fermi level
is pushed up at the same time. Even though the core excitation energy scale $E'$ is slightly decreased, the overall effect
 is dominated by the decrease of the scales $E$ and $\Delta E$ which favors the formation of the halo~\cite{rotival07a}.
\end{itemize}

By no means SLy4 or any other existing Skyrme parametrization of the nuclear EDF is to be seen as universal. As
exemplified above, it is found that the choice of the particle-hole Skyrme parametrization can affect
significantly the neutron canonical spectrum close to the Fermi energy. As a result, the drip-line position and
the halo energy scales are modified, hence the values of $N_{\mathrm{halo}}$ and $\delta R_{\mathrm{halo}}$. Scanning the set of
parameterizations used in the present study, $N_{\mathrm{halo}}$ and $\delta R_{\mathrm{halo}}$ can change by as much as $100\,\%$.
Thus, solid experimental data on medium-mass halo nuclei will be useful in the far future to constrain some characteristics of
the particle-hole part of the EDF. Note that this is at variance with the conclusions drawn in
\sect\ref{sec:pair_loc} regarding the pairing part of the energy functional.

Eventually, (reasonable) variations in the characteristics of currently used EDFs translate into large uncertainties in
the prediction of halo properties. For instance, no halo nucleus would be predicted at the drip-line of Cr
isotopes using the Skyrme \mbox{SkM$^\ast$} parametrization, as the semi-magic nucleus \mbox{$^{74}$Cr} is
predicted to be the last bound Cr isotope in this case. The same care is to be considered regarding the
large-scale prediction of spherical halo candidates over the nuclear chart presented in the next section.

\subsection{Spherical medium and heavy mass nuclei}
\label{sec:table}

Systematic predictions of halo properties from the \{SLy4+REG-V\} EDF are now presented. We restrict ourselves to
even-even spherical nuclei, as predicted by the Gogny D1S interaction~\cite{hilaire07}. Among all even-even nuclei, we
define the sub-set of "spherical" nuclei as those fulfilling the condition

\begin{equation}
|\beta_2|\equiv  \sqrt{\frac{5}{16 \pi}} \, \frac{4 \pi |Q_{2}|}{3 R^2 A}<0.1 \, \, \, ,
\end{equation}
where $Q_{2}$ is the axial mass quadrupole moment. The rather large interval allowed on
$\beta_{2}$ is arbitrary and does not distinguish between soft and rigid ground states. Such a condition provides
a list of about $500$ nuclei, in agreement with similar predictions made using Skyrme
functionals~\cite{doba04b}, although the drip-line positions slightly differ in the two models.

\begin{figure*}
\centering
\includegraphics[keepaspectratio,angle = -90, width = 0.85 \textwidth]%
{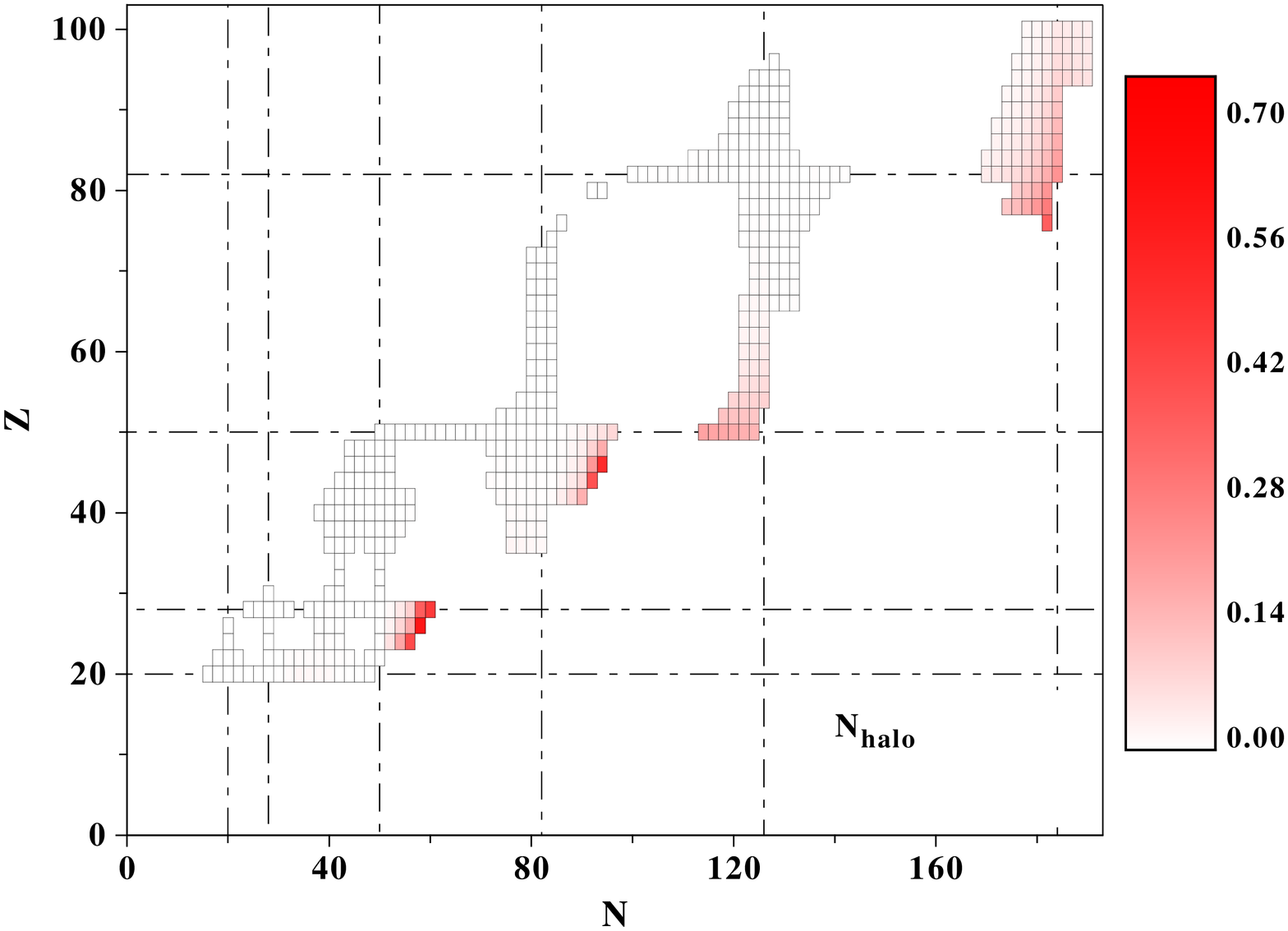} \caption{(Color Online) $N_{\mathrm{halo}}$ parameter predicted by the \{SLy4+REG-V\}
EDF for about $500$ spherical nuclei. \label{fig:syst_nhalo}}
\end{figure*}

\begin{figure*}
\centering
\includegraphics[keepaspectratio,angle = -90, width = 0.85\textwidth]%
{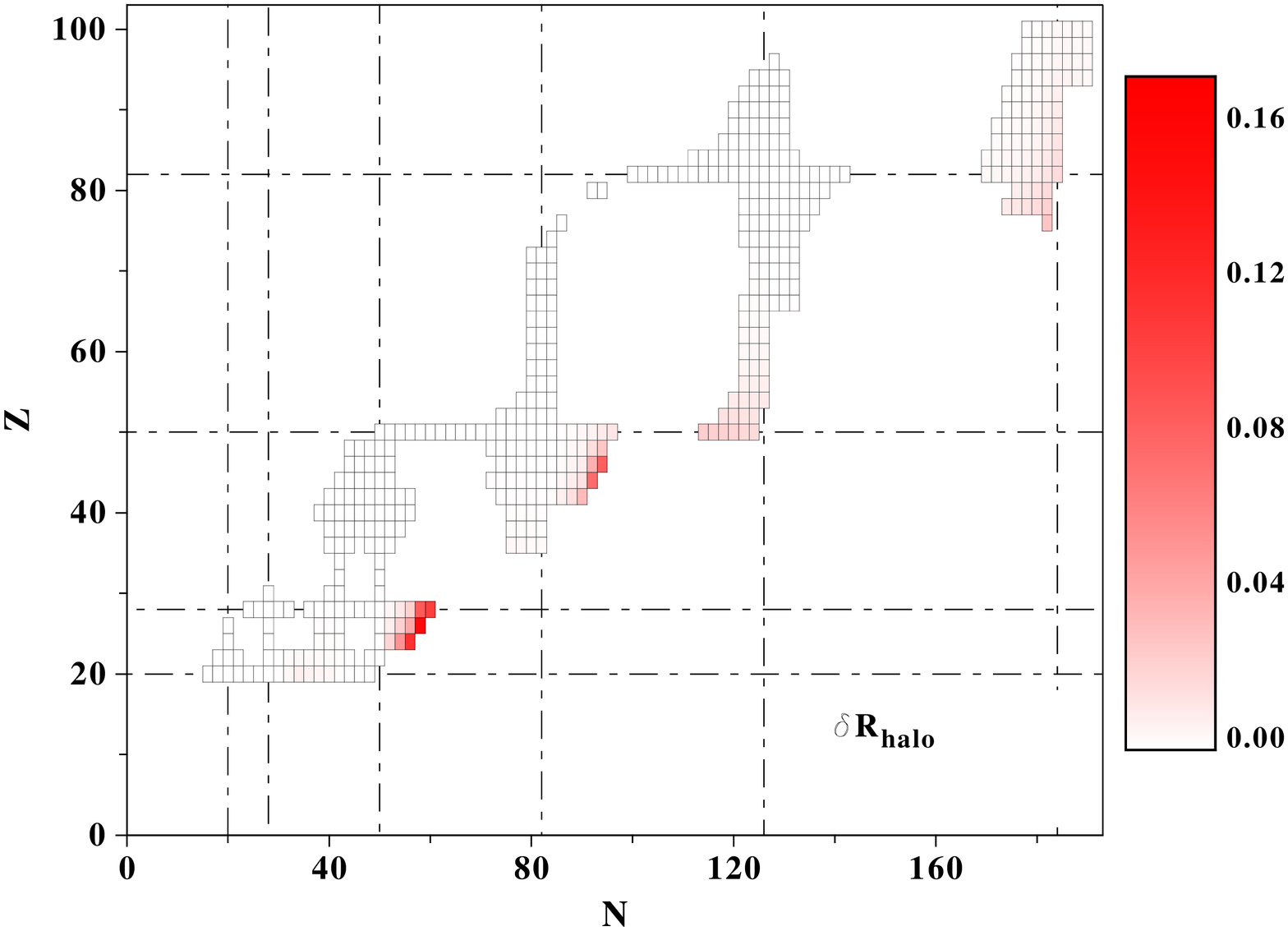} \caption{(Color Online) $\delta R_{\mathrm{halo}}$ parameter predicted by the
\{SLy4+REG-V\} EDF for about $500$ spherical nuclei.} \label{fig:syst_drhalo}
\end{figure*}

\begin{figure}[hptb]
\includegraphics[keepaspectratio, width=\columnwidth]{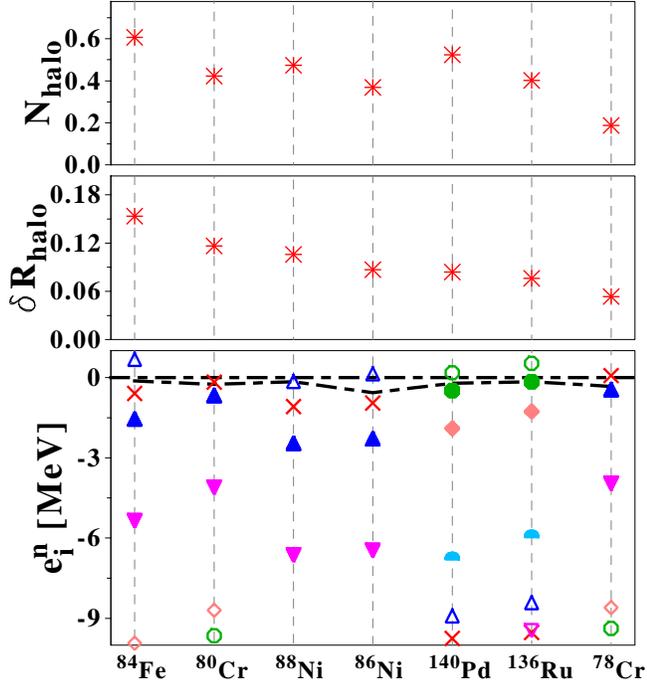}
\caption{(Color Online) Canonical spectrum and halo factors of the best halo candidates, as predicted with the \{SLy4+REG-V\}
functional. Conventions
for labeling individual states are given in \figu\ref{fig:ref}. \label{fig:best10}}
\end{figure}

\begin{figure}[hptb]
\includegraphics[keepaspectratio,angle = -90, width = \columnwidth]%
{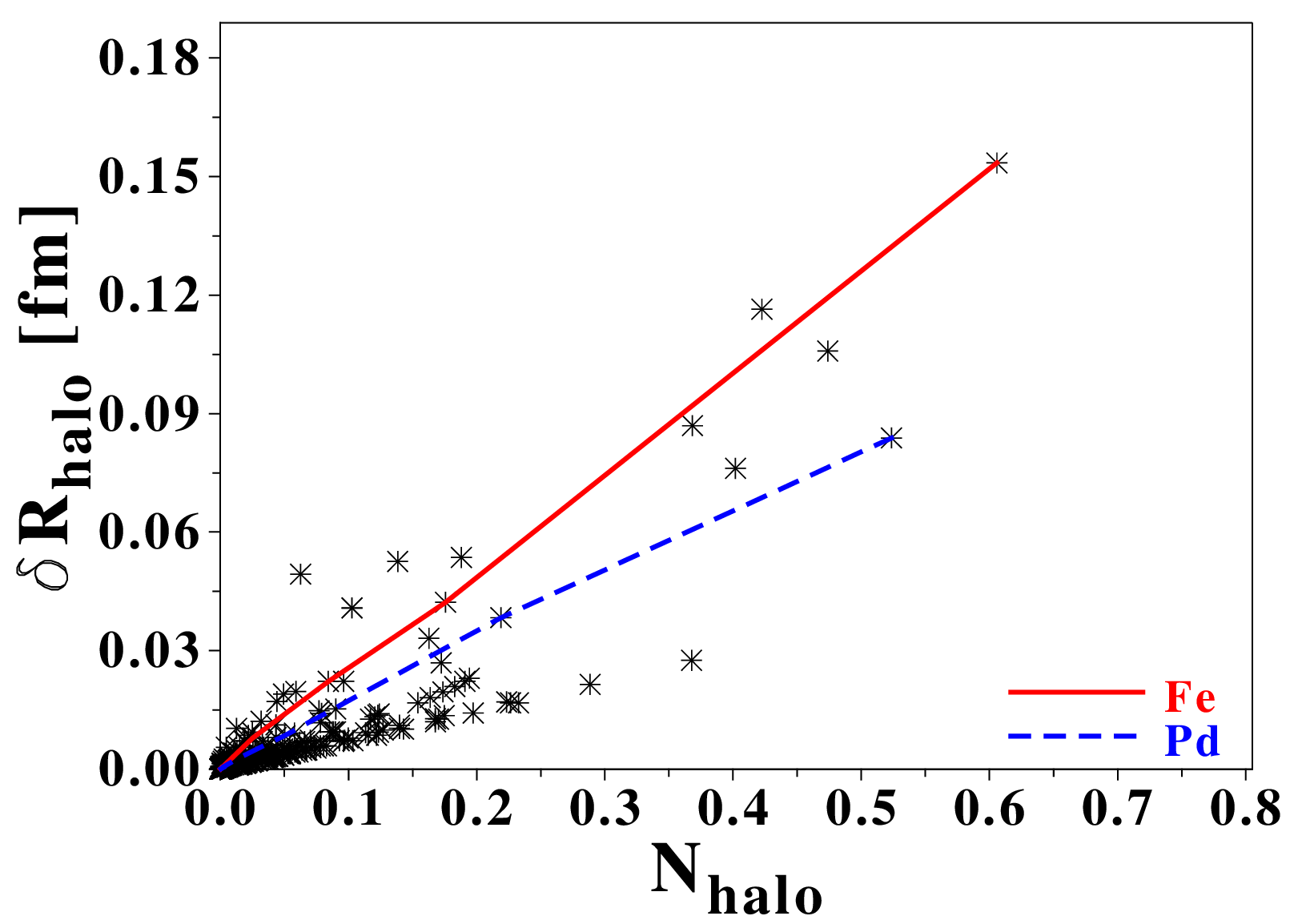} \caption{(Color Online) Correlation between $N_{\mathrm{halo}}$ and $\delta R_{\mathrm{halo}}$ for
all spherical nuclei computed with the \{SLy4+REG-V\} functional. The evolution of $\delta R_{\mathrm{halo}}$ as a function
of $N_{\mathrm{halo}}$ for Fe and Pd isotopes are highlighted.\label{fig:phase}}
\end{figure}

Results for $N_{\mathrm{halo}}$ are shown in \figu\ref{fig:syst_nhalo}. We observe that (i) several isotopic chains are
predicted to display neutrons halos, (ii) halos are predicted to exist only at the very limit of neutron stability,
(iii) the maximum value of $N_{\mathrm{halo}}$ is about $\sim0.7$, (iv) very few heavy elements in the (Pt, Hg, Tl...)
region are found to have a non-zero halo parameter $N_{\mathrm{halo}}$, (v) on the large scale, the halo phenomenon is very
rare and almost accidental, (vi) looking at the best cases between $Z=20$ to $Z=100$, the absolute and relative
values of $N_{\mathrm{halo}}$ decrease with nuclear mass.

Results for $\delta R_{\mathrm{halo}}$ are presented in \figu\ref{fig:syst_drhalo} and confirm the above analysis on
$N_{\mathrm{halo}}$. In particular, it is seen that the fraction of decorrelated nucleon has almost no influence on the
nuclear extension of massive nuclei. Only two very localized regions where the predicted halo significantly
affects the neutron r.m.s. radius are found, e.g. for (i) Cr, Fe and Ni nuclei, and (ii) Pd and Ru isotopes. The drip-line
isotopes of these elements are predicted as the best halo candidates for the \{SLy4+REG-V\} EDF.

An analysis of single-particle properties of the best halo candidates is found in \figu\ref{fig:best10}. These
nuclei have in common the presence of very weakly bound $s$ or $p$ states. Although states with larger angular
momentum contribute to the nuclear halo in some cases~\cite{rotival07a}, the presence of weakly bound
\mbox{$\ell=0,1$} orbitals remains mandatory for a significant halo to develop. That being said, no \mbox{pure}
$s$-wave or $p$-wave halo has been found, which demonstrates the collectivity of the phenomenon in medium-mass
systems.

The complementarity between the two criteria $N_{\mathrm{halo}}$ and $\delta R_{\mathrm{halo}}$ appears more clearly through the
large scale analysis. The plot presented in \figu\ref{fig:phase} shows that the two observables are correlated
within a given isotopic chains and the information carried by both quantities is somewhat redundant in this case. On a larger scale however, the
correlation pattern changes as the proton number increases (between Cr and Pd isotopes for instance), i.e.
$\delta R_{\mathrm{halo}}$ increases much less with $N_{\mathrm{halo}}$ with mass.

Finally, one can turn to the particular case of Zr and Ca isotopes which have been predicted to be "giant halo"
nuclei ~\cite{meng98,sandulescu03,geng04,kaushik05,grasso06,terasaki06}. In the present study, the values of
$N_{\mathrm{halo}}$ and $\delta R_{\mathrm{halo}}$ do not lead to such a conclusion. The first reason resides in the different
single-particle spectra predicted by the functional used in the present study. Considering
that relativistic models (TM1~\cite{zhang03}, NLSH~\cite{sharma93}, NL3~\cite{sandulescu03}...)
or non-relativistic ones
(SkI4~\cite{reinhard95}...) predict weakly
bound $p$ states~\cite{sandulescu03,geng04,kaushik05,grasso06,terasaki06} at the neutron drip-line of those
elements, it is likely that the application of our criteria on those results would lead to predicting the
existence of halos. However, we put into question the very notion of "giant halo"
that comes from summing up the occupations of weakly bound orbitals. Such a counting
procedure is qualitatively incorrect as
nucleons occupying weakly bound orbitals are mostly located within the nuclear volume.
Such an unjustified counting is a reminiscence of the denomination of ``1(2)-neutron'' halo used for light systems. However, the latter
relates to the fact that these light nuclei are well described by a ``core+1(2) neutrons''
cluster model whereas
practitioners are well aware that only a fraction of nucleon resides in average beyond the classical turning point. We prove in the present work that the fraction of neutron that participate in the halo does not scale with $A$ and remains in medium-mass nuclei of the same order as in light nuclei. The notion of spatial
decorrelation is key to the meaningful definition of the halo region and the halo parameters $N_{\mathrm{halo}}$ and
$\delta R_{\mathrm{halo}}$.

\section{Conclusions}
\label{sec:conclusion}

In Ref.~\cite{rotival07a}, a model-independent analysis method was developed to characterize halos in finite many-fermion systems. Eventually, the method leads to quantifying potential halos in terms of (i) the average number of
 fermions participating to it, (ii) its impact on the system extension.

In the present paper, the versatility of the method is illustrated by applying it to several quantum mesoscopic
systems, that is (i) few-body nuclei calculated through coupled-channels methods, (ii) atom-positron and ion-positronium complexes evaluated
through the fixed-core stochastic variational method and (iii) medium-mass nuclei calculated using the energy density
functional (EDF) method. The ability of the analysis method to quantitatively characterize halos of very different
scales underlines the universal nature of the halo phenomenon, namely the quantum tunneling into the classically forbidden region.

In the second part of the paper, the need for a coherent treatment of the ultra-violet divergence and of
the density dependence of the local pairing functional employed in EDF calculations is highlighted. Using a naive regularization scheme in combination
with a strongly surface-enhanced pairing functional leads ultimately to the formation of a uniform gas of di-neutrons dripping out of nuclei. When the pairing functional is such that this spurious gas does not completely develop (most probably for numerical reasons), it can be mistakenly interpreted as a halo. We point out that using a renormalization scheme or constraining the pairing functional through ab-initio calculations of pairing gaps in infinite nuclear matter prevents such problems from happening.

In the third part of the present paper, the impact of pairing correlations on halo systems is further studied.
First, it is shown that the pairing anti-halo effect might be counter-balanced by the pair scattering to less
bound orbitals, possibly with small orbital angular momenta. The net effect of pairing on the formation of halos greatly
depends on the detail of the single-particle spectrum of the nucleus under study. Second, \mbox{low-$\ell$}
orbitals are found to be less paired than \mbox{high-$\ell$} ones, but not enough to really decorrelate from the
pairing field, as it is the case for (non-realistic) extreme conditions~\cite{hamamoto03,hamamoto04}. Third, the
low-density-dependence of the pairing functional is shown to have almost no effect on the formation and characteristics of halos, as
long as it is combined with the renormalization scheme or constrained from ab-initio calculations, as mentioned above.

On the contrary, halo properties significantly depend on the characteristics of the particle-hole part of the
nuclear EDF. Thus, medium-mass halo systems might be more useful to constrain the particle-hole part of
the functional than its particle-particle counterpart. Unfortunately, such experimental data are not likely to
become available any time soon, even with the next generation of radioactive beam facilities. Indeed, although the
neutron density becomes accessible through several experimental
techniques~\cite{lubinski94,lubinski98,horowitz01,krasznahorkay04,kienle04,yako06}, extracting it in medium-mass
drip-line nuclei is more than a challenge for the decade(s) to come. In addition, the neutron r.m.s. radius is not
sufficient to study halos quantitatively, and other probes have to be envisioned~\cite{jensen04}. Of course, the
precise determination of the neutron r.m.s. radius and associated neutron skin in non-halo systems is already
crucial as it provides constraints on the physics of neutron stars~\cite{lattimer04,steiner05} and on the nuclear
symmetry energy~\cite{danielewicz03,yoshida06}, for instance.
As a result, one should focus at first on the study of neutron
skins in non-halo systems to constrain the isovector nature of the nuclear EDF. The fine tuning provided by
extreme exotic systems such as medium-mass halo nuclei will only come as a second step.

Still, it is of theoretical interest to understand the structure of halo nuclei and to assess their occurrence
over the nuclear chart. With that in mind, we performed large-scale calculations over all (predicted) spherical
even-even nuclei. It was concluded that (i) several isotopic chains may display neutron halos, (ii) halos can only
exist at the {\it very} limit of neutron stability, (iii) very few heavy elements, in the (Pt, Hg, Tl...) region,
are found as possible halo candidates, (iv) on the large scale, the halo phenomenon is very rare
and requires the presence of a low-lying state with an orbital angular momentum $\ell \leq 1$ at the drip line, (v) medium-mass
halos are more collective than in light nuclei, with non-negligible contributions from several low-lying states, including orbitals with  $\ell \geq 2$, (vi) the impact of the halo on the nuclear extension decreases with the increasing mass.

Let us now turn to potential works to come. First, several extensions of the newly proposed
method can be envisioned. For instance, the extraction of halo properties in deformed systems, which represent the
majority of known or predicted nuclei, requires additional formal developments, starting from a decomposition of
the nuclear density in multipolar moments
\begin{equation}
\rho(\vec{r}\,)=\sum_{\ell,m_\ell}\rho_\ell^{m_\ell}(r)
\,Y_\ell^{m_\ell}(\hat{r}\,)\,.
\end{equation}

The analysis regarding the relative asymptotic positioning of spectroscopic amplitudes and the occurrence of crossing patterns should thus be adapted to multipolar moments \mbox{$\rho_\ell^{m_\ell}(r)$}. Effects such as the increased contamination of weakly-bound
deformed orbitals by \mbox{$\ell=0$} components at the limit of stability, favoring the formation of halos in
deformed systems~\cite{hamamoto04b,hamamoto05,yoshida05,hamamoto06}, could be investigated. In addition to
deformed systems, the method should be extended to odd-even and odd-odd nuclei. This requires to formulate
the method for non-zero spin states. Also, the effects of an explicit treatment of long-range correlations, e.g.
symmetry restorations and large-amplitude collective motion, on medium-mass halo nuclei should be studied in
connection with the analysis method proposed in the present work. The explicit inclusion of such
correlations would allow a more reliable extraction of $N_{\mathrm{halo}}$ and $\delta R_{\mathrm{halo}}$ in light systems through
EDF calculations. For instance, the study of \mbox{$^{22}$C}, which is predicted to be a halo system by few-body
calculations~\cite{horiuchi06}, could provide a bridge between few- and many-body techniques.

Finally, it would be of interest to conduct the same study as the one presented in the present work using the
Gogny effective interaction to further probe the dependence of the results on the characteristics of the energy
density functional; e.g. finite-range effects. Such a study is not feasible using traditional codes making use of
a harmonic oscillator basis because the asymptotic of the one-body density is not described correctly enough to
extract $N_{\mathrm{halo}}$ and $\delta R_{\mathrm{halo}}$ reliably. However, a newly developed code expanding the quasiparticle
states on the eigenstates of a Woods-Saxon potential should allow one to do so~\cite{schunck07a}.

\section*{Acknowledgments}

Work by K.~B. was performed within the framework of the espace de structure nucl{\'e}aire th{\'e}orique (ESNT).
T.~D. acknowledges the hospitality of the SPhN and ESNT on many occasions during the elaboration of this work. V.
R. wishes to thank the NSCL for its hospitality and support. We wish to thank
J.~Mitroy for providing us with his results on atom-positron complexes and for fruitful discussions, F.~Nunes for
her valuable input on few-body models and halos, T.~Lesinski for providing us with the parameterizations
$\rho^{1/2/3}_{\mathrm{sat}}$ and $m^\ast 1$, as well as A.~S.~Jensen and K.~Riisager for very instructive conversations. This work
was supported by the U.S. National Science Foundation under Grant No. PHY-0456903.

\bibliography{bib-data}

\end{document}